\documentclass[fleqn,usenatbib,twocolumn,twocolappendix]{openjournal}

\usepackage{newtxtext,newtxmath}

\usepackage[T1]{fontenc}
\usepackage{ae,aecompl}
\usepackage{lsstdesc_macros}

\usepackage[colorlinks=true, allcolors=blue]{hyperref}

\usepackage{graphicx}	
\usepackage{epstopdf}
\usepackage{amsmath}	
\usepackage{amssymb}	
\usepackage{float}
\usepackage{multirow,array}
\usepackage{soul}
\usepackage[usenames, dvipsnames]{color}  
\usepackage[usenames, dvipsnames]{xcolor}  
\usepackage{comment}
\usepackage[caption=false]{subfig}
\usepackage{soul}
\usepackage{hyperref}
\usepackage{lineno}

\setstcolor{red}

\defcitealias{Schmidt20}{Schmidt \& Malz et al.}

\newcommand{\citeDCt}{\citetalias{Schmidt20} (\citeyear{Schmidt20})\xspace}

\newcommand{\orcidauthor}[3]{\author{\href{http://orcid.org/#1}{#2$^{#3}$}}}

\shorttitle{DeepDISC photo-z}

\begin{document}

\title[DeepDISC: LSST photo-z]{DeepDISC-photoz: Deep Learning-Based Photometric Redshift Estimation for Rubin LSST}

\orcidauthor{0009-0005-7923-054X}{Grant Merz}{1,*}
\orcidauthor{0000-0003-0049-5210}{Xin Liu}{1,2,3, \dagger}
\orcidauthor{0000-0002-5091-0470}{Samuel Schmidt}{4}
\orcidauthor{0000-0002-8676-1622}{Alex I. Malz}{5,6}
\orcidauthor{0000-0002-5596-198X}{Tianqing Zhang}{5,7,8}
\orcidauthor{0009-0009-7822-7110}{Doug Branton}{5,9}
\orcidauthor{0000-0001-9947-6911}{Colin J. Burke}{10}
\orcidauthor{0000-0002-1074-2900}{Melissa Delucchi}{5,6}
\orcidauthor{0009-0001-2278-8199}{Yaswant Sai Ejjagiri}{11}
\orcidauthor{0009-0009-2281-7031}{Jeremy Kubica}{5,6}
\orcidauthor{0000-0003-4247-0169}{Yichen Liu}{1,2,12} 
\orcidauthor{0000-0001-5028-146X}{Olivia Lynn}{5,6}
\orcidauthor{0000-0001-6984-8411}{Drew Oldag}{5,9}
\author{the LSST Dark Energy Science Collaboration}

\thanks{$^*$ Corresponding Author: \href{mailto:gmerz3@illinois.edu}{gmerz3@illinois.edu}.}
\thanks{$^\dagger$ Corresponding Author: \href{mailto:xinliuxl@illinois.edu}{xinliuxl@illinois.edu}.}

\affiliation{$^{1}$Department of Astronomy, University of Illinois at Urbana-Champaign, 1002 West Green Street, Urbana, IL 61801, USA}
\affiliation{$^{2}$National Center for Supercomputing Applications, University of Illinois at Urbana-Champaign, 1205 West Clark Street, Urbana, IL 61801, USA }
\affiliation{$^{3}$Center for Artificial Intelligence Innovation, University of Illinois at Urbana-Champaign, 1205 West Clark Street, Urbana, IL 61801, USA }
\affiliation{$^{4}$Department of Physics and Astronomy, University of California, Davis, CA, 95616, USA}
\affiliation{$^{5}$LSST Interdisciplinary Network for Collaboration and Computing Frameworks, 933 N. Cherry Avenue, Tucson AZ 85721 }
\affiliation{$^{6}$The McWilliams Center for Cosmology \& Astrophysics, Department of Physics, Carnegie Mellon University, Pittsburgh, PA 15213, USA }
\affiliation{$^{7}$Department of Physics and Astronomy and PITT PACC, University of Pittsburgh, Pittsburgh, PA 15260, USA}
\affiliation{$^{8}$SLAC National Accelerator Laboratory, 2575 Sand Hill Road, Menlo Park, CA 94025, USA}
\affiliation{$^{9}$DiRAC Institute and the Department of Astronomy, University of Washington, 3910 15th Ave NE, Seattle, WA 98195, USA }
\affiliation{$^{10}$Department of Astronomy, Yale University, 266 Whitney Avenue, New Haven, CT 06511, USA}
\affiliation{$^{11}$Department of Computer Science, University of Illinois at Urbana-Champaign, 201 North Goodwin Avenue, Urbana, IL 61801, USA }
\affiliation{$^{12}$Steward Observatory, University of Arizona, 933 N Cherry Ave, Tucson, AZ 85719}




\begin{abstract}
Photometric redshifts will be a key data product for the Rubin Observatory Legacy Survey of Space and Time (LSST) as well as for future ground and space-based surveys.  The need for photometric redshifts, or photo-zs, arises from sparse spectroscopic coverage of observed galaxies.  LSST is expected to observe billions of objects, making it crucial to have a photo-z estimator that is accurate and efficient.  To that end, we present \textsc{DeepDISC photo-z}, a photo-z estimator that is an extension of the \textsc{DeepDISC} framework.  The base DeepDISC network simultaneously detects, segments, and classifies objects in multi-band coadded images.  We introduce photo-z capabilities to \textsc{DeepDISC} by adding a redshift estimation Region of Interest head, which produces a photo-z probability distribution function for each detected object.  On simulated LSST images, \textsc{DeepDISC} photo-z outperforms traditional catalog-based estimators, in both point estimate and probabilistic metrics.  We validate \textsc{DeepDISC} by examining dependencies on systematics including galactic extinction, blending and PSF effects. We also examine the impact of the data quality and the size of the training set and model.  We find that the biggest factor in \textsc{DeepDISC} photo-z quality is the signal-to-noise of the imaging data, and see a reduction in photo-z scatter approximately proportional to the image data signal-to-noise.  Our code is fully public and integrated in the RAIL photo-z package for ease of use and comparison to other codes at \href{https://github.com/LSSTDESC/rail\_deepdisc}{https://github.com/LSSTDESC/rail\_deepdisc}.
\end{abstract}


\section{Introduction}

\label{sec:intro}

Accurate and reliable redshift measurements are of great importance in a multitude of astronomical analyses. Redshifts are necessary to convert observed angular sizes to distances, and needed in studies of weak and strong gravitational lensing \citep{Mandelbaum18, Treu10}, large scale structure \citep{Seo12,DESBAO22}, galaxy evolution \citep{Finkelstein15}, galaxy clusters \citep{Allen11, Wen21} and more \citep{Newman22}. The most reliable method of measuring an object's redshift comes from comparing the true and observed wavelengths of known spectral features.  However, spectroscopic follow-up is time consuming and expensive for large surveys. In the era of the Legacy Survey of Space and Time \citep{LSST19}, billions of objects will be observed, making even a 10\% complete spectroscopic catalog impossible.  Despite a lack of complete spectroscopy, redshifts can still be obtained using photometry, because the photometric flux or magnitude of an object in a given filter is equivalent to the underlying spectral energy distribution (SED) integrated over the filter transmission function.  Measured fluxes of an object thus represent a sparse sampling of the SED, and spectral features will appear as color differences in the source photometry, which can be used to estimate redshift. Photometric redshifts (photo-zs) provide a much more efficient alternative at the price of a lower precision and accuracy, enabling processing of large numbers of objects.  While efficient, photo-z methods face the inherent challenge of estimating an object's redshift from a low-dimensional set of photometric features. Degeneracies in the photometric feature space can cause catastrophic photo-z outliers \citep{Massarotti01}.  Given this limitation, it is common for these algorithms to produce a probability density function (PDF) for each redshift, rather than a single number.  This allows for a full characterization of the redshift uncertainty due to ambiguous spectral feature localization from the low SED resolution provided by the photometry. Photo-z PDFs can subsequently be integrated into Bayesian analyses \citep{Meyers09, DES16,  DES18, DES21, HSC23, Mitra23} and will be a necessary product for LSST \citep{LSST_SRD}.

Standard photometric redshift estimation methods largely fall into two categories: template-based SED fitting and machine learning algorithms \citep[See][for a review]{Salvato19}, although hybrid approaches have been implemented \citep{Tanigawa24}. Standard photo-z methods assume a pre-defined set of features; most commonly fluxes or magnitudes and colors that are derived from image processing pipelines are used as input. In contrast, there has been increasing focus on using images themselves rather than pre-computed features \citep{DIsanto18, Pasquet19, Dey22, Hayat21, Schuldt21, Treyer24, Roster24}.  Using pixel-level information incorporates morphological features such as size and color gradients across a source light profile.  Leveraging this information, feature extractors such as convolutional neural networks (CNNs) have been shown to outperform traditional photo-z methods.  In addition to new architectures, new training methodologies have been explored such as adversarial training \citep{Campagne20} and contrastive learning approaches that pre-train a model to extract features, and then use those features for downstream tasks, e.g., photo-z estimation \citep{Hayat21, Lanusse23}.    

While many image-based methods outperform traditional feature-based methods, there are still some inherent limitations.  Most photo-z studies employing machine learning methods rely on pre-classification of sources in order to avoid stellar contamination or to focus on a particular class of object \citep[\eg][]{Schuldt21}. However, \cite{DIsanto18} test an image-based photo-z estimator on a mixed sample of stars, quasars, and galaxies, and do not find a significant drop in performance due to contamination.  \cite{DIsanto18} recommend that future studies utilizing image-based methods adopt this methodology to test model performances in the face of contamination, but this has not been widely adopted by the community.  Stellar contamination strongly increases as fainter objects are observed, due to traditional morphological classification schemes that confuse stars for small point-source-like galaxies \citep{Bosch18}.  

Another problem that affects not just image-based methods is blending. Blending, or the visual overlap of sources in astronomical images, is a systematic problem that now plagues modern astronomical surveys.  Most blending is caused by the projection of sources existing in the same line of sight, with a minority of cases being physical mergers.  Increased observational depth causes more sources to be detected, which increases the observed rate of blending \citep{melchior_challenge_2021}. In the shallowest fields of the Hyper Suprime-Cam Subaru Strategic Program \citep[HSC SSP][]{aihara_hyper_2018}, blending already affects a majority of sources \citep{Bosch18}.  This systematic and can cause significant biases in weak lensing studies which rely on accurate shape measurements \citep{Mandelbaum18}. Blending also effects source photometry \citep{Boucaud19} which will naturally affect photo-z estimation.  With most image-based methods, all pixels in a cutout are used, meaning that blended objects are included in the automatic feature extraction of the networks \citep{Pasquet19, Dey22, Hayat21}. \cite{Schuldt21} employ a masking procedure to address blended companions, which they find increases their model performance.  Image-based estimators provide an advantage in their simultaneous treatment of blending and photo-z estimation, but typically rely on source catalogs produced by traditional detection and deblending pipelines for curating the input images.

In this work, we aim to address both blending and source classification with an image-based photo-z estimator \textsc{DeepDISC photo-z}.
\textsc{DeepDISC} \cite{Merz2023} is a framework for applying instance segmentation models to astronomical images and is derived from the method developed in \cite{Burke_deblending_2019}.  Instance segmentation models are designed to detect and segment objects in images, as well as perform downstream tasks such as classification.  Fundamentally, instance segmentation models are composed of a backbone network that learns to extract features from images, a Region Proposal Network which learns to detect and localize objects, and Region of Interest heads which learn to predict a downstream measurement for each detected object.  \cite{Merz2023} applied this framework to multi-band HSC imaging to detect, segment, and classify objects as stars or galaxies.  In their comparison study using different backbone networks, they found that vision transformer backbone networks \citep{Li21, Liu21} outperformed convolutional neural networks at detection, segmentation and classification of astronomical scenes.  \cite{Merz2023} used models pre-trained on ImageNet, a set of terrestrial RGB image data, and fine-tune on 3 filter (g, r and i) images with different contrast scalings.  As a supervised model, one of the main limitations of \textsc{DeepDISC} is that a deblended ground truth must be provided during training in order to detect and segment deblended objects during inference or testing. Biases in the ground truth (which cannot be known with real data) may propagate to the inferred output.

We present the development of the photo-z module of the \textsc{DeepDISC} framework, \textsc{DeepDISC-photoz}. It estimates probabilistic photometric redshifts directly from multi-band images, circumventing the need for feature extraction, pre-classification of sources, and the need for a separate detection step entirely. \textsc{DeepDISC} produces probabilistic redshift estimates, capturing meaningful uncertainties and degeneracies inherent in mapping redshifts from photometry, evident in the structure of its estimated PDFs.  The aim of this work is to compare \textsc{DeepDISC} photo-z to catalog-based methods by benchmarking on simulated data, and to investigate the performance of our model under different observing conditions and training methodologies.  We interface \textsc{DeepDISC} with an open-source library designed for end-to-end photo-z testing and place \textsc{DeepDISC} photo-z in the wider context of photo-z literature, with a focus on application to LSST data.

This paper is organized as follows. In \S\ref{sec:data}, we describe the simulated data and pre-processing steps. In \S\ref{sec:methods}, we describe our model and training methodology as well as the other photo-z estimators used in our analysis. We present the results of our trained photo-z code in \S\ref{sec:results}, and quantify our photo-z estimation performance using various metrics.  We discuss the results along with limitations and future applications in \S\ref{sec:discussion}.  In \S\ref{sec:conclusions}, we contextualize our findings and conclude.

\section{Data}
\label{sec:data}

\begin{figure}
    \centering
    \includegraphics[width=\columnwidth]{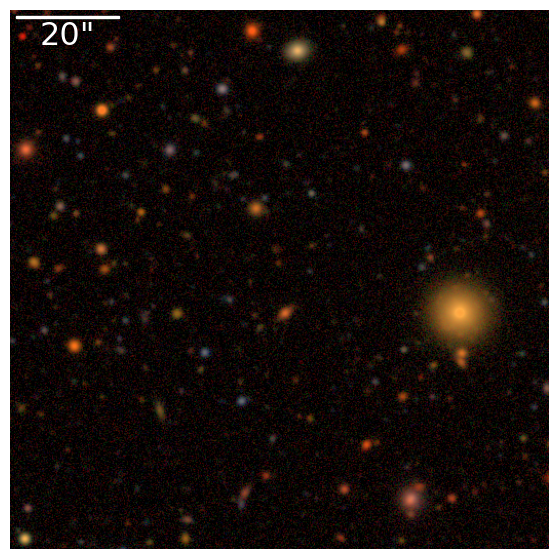}
    \caption{An example DC2 image used for training the network.  The RGB image corresponds to i, r and g bands and has been scaled with a Lupton asinh scaling \citep{Lupton2004} for visualization purposes.}
    \label{fig:image_ex}
\end{figure}

\begin{figure}
    \centering
    \includegraphics[width=\columnwidth]{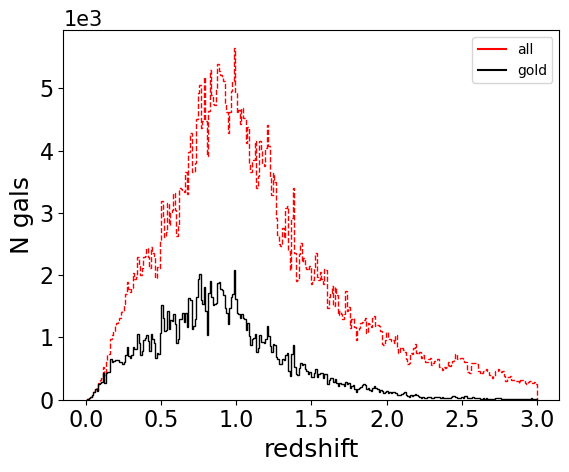}
    \caption{Histogram of the redshift distribution of objects in the training set. While the detection and segmentation branches of \textsc{DeepDISC} include all objects during training, the redshift estimation branch only trains using the magnitude-limited ``gold" sample.}
    \label{fig:trainz_hist}

\end{figure}

In this work, we utilize the second Data Challenge (DC2) simulated data from the LSST Dark Energy Science Collaboration (DESC). The simulated images and associated truth catalogs give a ground truth which allows us to perform a controlled experiment.  In using a representative data set we have ignored selection effects in spectroscopic samples.  The data production is described in \cite{DC2}. To briefly summarize, the cosmoDC2 extragalactic catalog \citep{CosmoDC2} is created starting with an N-body simulation of 10,240$^3$ particles in a volume of 4.225 Gpc$^3$. Simulations are repeated to obtain a volume sufficient for redshifts of $z\leq3$. Particle and halo lightcones are produced by interpolating snapshots in time. Galaxies are assigned to halos using the UniverseMachine synthetic galaxy catalog, which contains halos populated with sythetic galaxies.  Galaxies are assigned to cosmoDC2 halos based on UniverseMachine halos of similar mass, while preserving the conditional distributions of star formation rate, stellar mass, and halo occupation density. Galaxy properties are assigned based on a combination of empirical and semi-analytic modeling. These properties include SEDs, photometry, and morphology.  A bulge+disk model is used for each galaxy's morphology, with separate SEDs for each component.  Disk and bulge sizes are determined from a fit to the size-luminosity relation of Sloan Digital Sky Survey data from \cite{Zhang19}. Additionally, more realism is added to the galaxy morphologies with ``knots" of point sources which model star-forming regions.  Knots are assigned based on an empirical model fit to HST data and use the same SED as the disk.

The image simulation software, imSim, uses GalSim \citep{GalSim} and includes physical effects due to the atmosphere and models LSST sensor features such as brighter-fatter and tree-ring effects.  For a complete description, see \cite{DC2}.  Systematic effects such as blending and variations in observing conditions are present in the simulated images and propagate into aperture photometry measurements.  An example image is shown in Figure \ref{fig:image_ex}.

In this work, we validate \textsc{DeepDISC} by self-consistently comparing with other photo-z methods using this dataset, and investigate how the model performs under different observational systematics and methodological setups. While physically and empirically motivated, the simulated data we use here has certain limitations which prevent full realism.  Simple parametric bulge+disk+knot light profiles do not capture the full complexity of real galaxies, which may be much more useful for image-based estimators.  The size-redshift relations used in the cosmoDC2 catalogs may not fully capture the rich diversity of galaxy properties over cosmic time.  Additionally, due to computational limitations, there were a limited number of SEDs to assign to galaxies at high redshifts.  This produces some unrealistic discreteness in the color space at high redshifts, which may benefit machine learning methods.



We create a training and test set of images and ground truth information using the GCRCatalogs\footnote{\href{https://github.com/LSSTDESC/gcr-catalogs}{https://github.com/LSSTDESC/gcr-catalogs}} package and LSST Butler \citep{LSSTButler}.  We use coadded imaging data corresponding to 5 years of observing.  The training set consists of 1048 cutouts of 525 pixels square. We randomly sample cutouts from a set of 12 LSST tracts.  In order to train \textsc{DeepDISC}, we must provide the network with object locations, segmentation masks, and redshifts for objects in the training set of cutouts.  Since we wish to produce a deblended catalog of photo-zs during inference, the ground truth we provide during training must be deblended. We provide the network with deblended truth information for each training set cutout by querying the truth catalogs within each image's footprint, up to a magnitude cut corresponding to 5$\sigma$ above the sky background in any filter.  This magnitude cut is applied to train the network to detect objects with sufficient signal-to-noise.  Then, we produce deblended bounding boxes and segmentation masks for the training set by running the deblending code \texttt{scarlet} \citep{melchior_scarlet_2018} using the scheme outlined in the quick start tutorial\footnote{\href{https://pmelchior.github.io/scarlet/0-quickstart.html}{https://pmelchior.github.io/scarlet/0-quickstart.html}}. We provide the multi-band PSF image evaluated at the center of each cutout to \texttt{scarlet} to aid in deblending.

Using the method described above, each image contains roughly 500-700 true objects. Distributions of object redshifts  
are shown in Figure \ref{fig:trainz_hist}.  We further limit the sample for redshift estimation to the LSST ``gold'' sample, defined by sources with i-band magnitude < 25.3 mag.  In total, the redshift estimation branch of the network is trained with 161,205 objects. We set aside 48 images in the training set as a validation set.  The test set is composed of 1925 images of 1050 pixels square.  Once trained, \textsc{DeepDISC} can be applied to images of any size, provided enough compute resources. The details of the redshift estimation are described in the following section.

\section{Methods}
\label{sec:methods}

\begin{figure*}[htbp]
    \centering
    \subfloat{\includegraphics[width=0.8\textwidth]{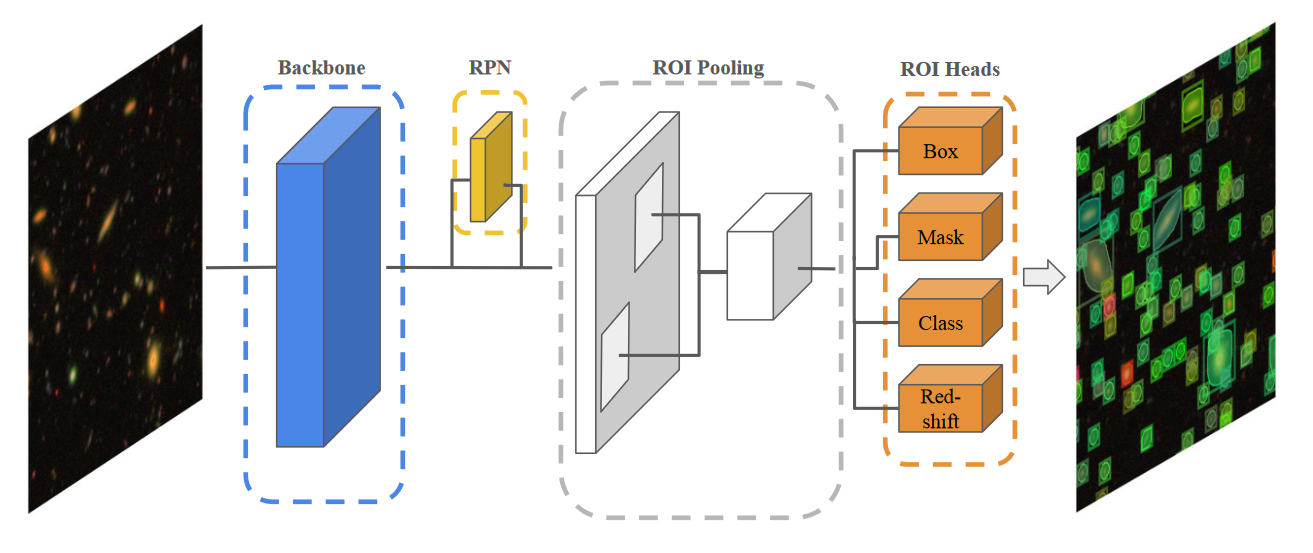}} \quad
    \subfloat{\includegraphics[width=0.8\textwidth]{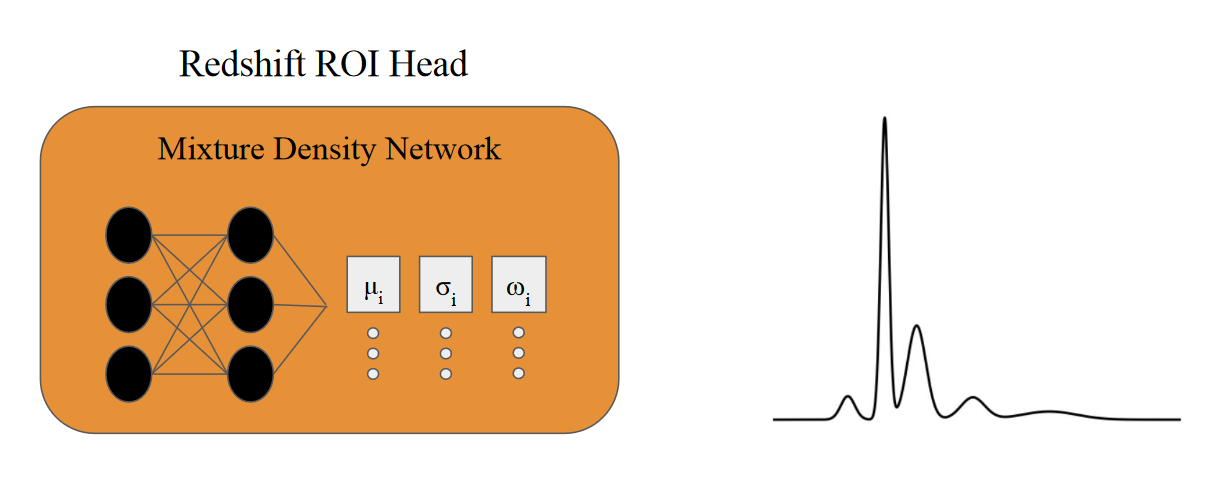}}
    \caption{Top: High-level diagram of the DeepDISC architecture. A multi-band image is input to the backbone network, which extracts features used for downstream tasks.  The Region Proposal Network (RPN) is trained to identify which parts of the image contain an object.  After these regions are identified, the corresponding features from the backbone are extracted and pooled together.  The Region of Interest (ROI) heads then perform the downstream tasks of bounding box and segmentation mask regression and object classification.  In this work, we add a redshift estimation ROI head.  Bottom: The redshift estimation is performed with a Mixture Density Network which will take the features from the previous stages and output a collection of weights, means, and standard deviations of a Gaussian mixture model which parametrizes the redshift PDF.}
    \label{fig:DDdiagram}
\end{figure*}

\textsc{DeepDISC}\footnote{\href{https://github.com/grantmerz/deepdisc}{https://github.com/grantmerz/deepdisc}} is implemented with PyTorch \citep{Pytorch} and the \texttt{detectron2} repository \citep{wu2019detectron2}. We choose as baseline backbone network a Multi-scale Vision Transformer \citep[MViTv2;][]{Li21} with a feature pyramid architecture \citep{FPN} and cascade Region of Interest (ROI) heads \citep{Cai17}.  The feature pyramid network extracts features from different resolution scales in the input image, allowing for increased flexibility with objects of different sizes. Features are then sent to the Region Proposal Network (RPN) to predict where objects are in the image, and then the features corresponding to each region are pooled and sent to the ROI heads.  A cascade ROI head iterates the detection stage at increasing confidence thresholds in order to produce higher quality detection inferences. A diagram of the \textsc{DeepDISC} architecture is shown in Figure \ref{fig:DDdiagram}.

Object detections are handled by the RPN and ROI heads.  The RPN proposes a set of bounding boxes as object detections.  Proposed bounding boxes are matched to the ground truth bounding boxes using an intersection-over-union (IOU) threshold.  Perfectly overlapping detected/truth boxes yield an IOU of 1, those with no overlap at all yield an IOU of 0. We use the \texttt{detectron2} standard IOU thresholds of [0.3,0.7], where predicted objects with a ground truth IOU less than 0.3 are considered "background", i.e., not a detection, those with 0.3<IOU<0.7 are ignored, and those with IOU>0.7 are considered to be a positive detection.  Highly blended objects may result in a mismatched detection during training if the IOU criterion is met, but the goal of the RPN and ROI heads are to refine these detections during training, with the provided truth information.  For more details on the RPN detection process, we refer the reader to \cite{Ren2015} and \cite{he_mask_2018}.

In order to produce a redshift estimate for each object, we add a Mixture Density Network \citep[MDN][]{bishop1994mixture} to the ROI heads. This allows the network to produce a PDF for each object's redshift, by using a Gaussian mixture model parametrization. \cite{DIsanto18} use a convolutional neural network and a MDN to produce photo-z PDFs for galaxy and quasar images in SDSS, and find that 5 Gaussians in the MDN work best.  We adopt this value for our MDN, after tests with 3 and 7 showed no significant improvements. The output of the MDN is a vector of length 15, where the first 5 components are the weights, the next 5 are the means, and the last 5 are the log of the standard deviations of the Gaussian components.  We  use the negative log likelihood as the loss function in the redshift branch of the network.  During training, all ground-truth objects are input into the network and used for detection and segmentation, but we only supply objects with an i-band magnitude < 25.3 mag to the redshift ROI head.  This magnitude-limited "gold" sample contains objects with a high signal-to-noise ratio and will be used for cosmological studies for LSST \citep{LSST_SRD}.  We include the reddening due to Milky Way dust as an extra neuron input to the MDN, by using the dust map from \cite{Schlafly11} provided by the \texttt{dustmaps} \citep{Dustmaps} package to get the E(B-V) value at the position of the objects in the training set.
\begin{table}
    \centering
    \begin{tabular}{ccc}
        \hline
        \hline
         & shape & activation  \\ 
         \hline
         Layer 1 &  (12544 + 1, 1024) & tanh  \\
         Layer 2 &  (1024, 64) & tanh \\
         Layer 3 &  (64, 15) &  -- \\
         \hline
    \end{tabular}
    \centering
    \caption{Number of layers and neurons in the MDN.  After features have been extracted and localized, they are flattened into a 12544 vector for each object.  After the dust extinction is concatenated to this vector, it is input to the MDN, which outputs the means, standard deviations, and normalized weights for 5 Gaussians. }
    \label{tab:MDN}
\end{table}

We adopt the same transfer learning strategy as in \cite{Merz2023} by using a MViTv2 backbone \citep{Li21} with weights that have been initialized from pre-training on the ImageNet1k \citep{ImageNet} dataset of terrestrial images.  Pre-training is often employed to speed up network convergence and reduce training costs.  While the ImageNet data is very different from astronomical scenes,  \cite{Burke_deblending_2019} and \cite{Merz2023} show that transfer learning is a viable strategy in training instance segmentation models to infer object properties in simulated and real astronomical data.  We discuss the implications and potential drawbacks of pre-training with ImageNet more in Section \ref{sec:results}.

\cite{Burke_deblending_2019} and \cite{Merz2023} used astronomical images with 3 filters, testing knowledge transfer from the RGB to another 3-color space. Photo-z estimation should use all available filters to sample object SEDs across the widest range of wavelengths possible.  In this work, we use all available filters for DC2 (\textit{ugrizy}) and thus test the viability of transfer learning to new color domains. In principle, DeepDISC can be applied to data with different filter sets, which in practice amounts to changing a single user-defined parameter before training.

Our choice of MViTv2 backbone network is motivated by the comparison study in \cite{Merz2023}.  This backbone produced the best results for object detection, deblending, and star-galaxy classification, and was more robust to contrast scalings. Contrast scaling is a commonly employed strategy to reduce the dynamic range of images input to a neural network \citep{Gonzalez2018}.  Astronomical images can have dynamic ranges that span several orders of magnitude, and so common strategies are to normalize images so that pixels fall in the range [0,1] and/or apply scaling functions such as an asinh scaling \citep{Lupton2004} to ensure that faint extended regions of objects are not ignored due to very bright central regions.  \cite{Merz2023} found the MViTv2 backbone to be very robust to different contrast scalings for instance segmentation tasks. Networks trained with scalings with large or small dynamic ranges showed little difference in performance.  Based on those results, we apply no scaling to the DC2 data and use the raw images.  This poses the advantage of maintaining noise and background properties that may otherwise be altered through changing the dynamic range.

For our initial training steps, we set the learning rate to 0.001 and train the entire network without fine-tuning, i.e., no freezing of the backbone weights. We train for 50 total epochs, lowering the learning rate by a factor of 10 each time at 15, 25, and 35 epochs.  A common strategy in transfer learning is to fine-tune, or freeze the deep layers of the network and only train the first few layers on the new data. Fine-tuning is commonly used in studies where the number of channels (colors) in the dataset used in the pre-training phase matches the number of dataset channels in the full training phase. However, the 6-channel \textit{ugrizy} astronomical images we use to train the network are very different from the ImageNet RGB images of everyday objects and scenes.  The MViTv2 backbone must learn to extract this information to help the rest of the network. Therefore, we do not fine-tune, and instead let all layers of the network learn during training.  It takes 6.5 hours to train \textsc{DeepDISC} on 4 NVIDIA V100 GPUs using our training set of images.


\subsection{RAIL integration}
In order to interface \textsc{DeepDISC} to the existing DESC photo-z ecosystem, we present \textsc{rail\_deepdisc}\footnote{\href{https://github.com/LSSTDESC/rail_deepdisc}{https://github.com/LSSTDESC/rail\_deepdisc}}, an open-source repo that interfaces \textsc{DeepDISC} with the DESC software Redshift Assessment Infrastructure Layers \citep{RAIL}.  \textsc{RAIL} is designed for end-to-end photo-z pipeline testing, with modules for creating data, applying photo-z estimators and evaluating their performance.  We design \textsc{rail\_deepdisc} to live under the RAIL ecosystem, meaning that once rail is installed, a user can import \textsc{rail\_deepdisc} as an optional dependency along with other photo-z estimators. \textsc{rail\_deepdisc} includes a streamlined configuration file API, interfaces with existing RAIL code, and fully parallel training and testing capabilities.  Both \textsc{rail\_deepdisc} and \textsc{DeepDISC} have been integrated with the LINCC Frameworks Python Project Template \cite{Oldag_2024}, which includes continuous integration tools for code tests, coverage, documentation, and package installation. We encourage the use of  \textsc{rail\_deepdisc} as a guide to implement other image-based codes within \textsc{RAIL}. 

\subsection{Code Comparison}
\label{subsec:cat-codes}

Hereafter, we use the terms "train" and "inform" interchangeably to match the RAIL convention. After training, we evaluate the performance of \textsc{DeepDISC} and compare to two catalog-based codes: Bayseian Photometric Redshifts \citep[BPZ;][]{BPZ} and \textsc{FlexZBoost} \citep[FZB;][]{FZB1, FZB2}.  Both codes exist as RAIL packages \footnote{\href{https://github.com/LSSTDESC/rail_bpz}{https://github.com/LSSTDESC/rail\_bpz}}\textsuperscript{,}\footnote{\href{https://github.com/LSSTDESC/rail_flexzboost/}{https://github.com/LSSTDESC/rail\_flexzboost/}} and were among the most competitive template-based and machine learning codes in the comparison study conducted by \citeDCt on simulated LSST data.

Both catalog-level codes are informed by the same training data set. In order to conduct a fair comparison with \textsc{DeepDISC}, we take the set of 161,205 gold sample entries in the \texttt{truth} catalog used by \textsc{DeepDISC} and take the corresponding matched set of entries in the \texttt{object} catalog.  The \texttt{object} catalog is produced by running the LSST Science pipelines software on the simulated images, which detects and deblends objects \citep{DC2}. We train BPZ and FZB using the \texttt{object} catalog as this incorporates blending and other imaging systematics, which \textsc{DeepDISC} encounters at the image level. This ensures all methods are informed with as close to the same prior information as possible.  However, we note that due to a flaw in the dust extinction law assumed in the simulations, the colors of our simulated galaxies are not fully represented by the template set employed by BPZ. This mismatch, along with the presence of imaging systematics in the derived photometric catalog will very likely degrade the performance of BPZ relative to \cite{Schmidt20}.

In order to assign redshifts to the \texttt{object} catalog to use for training, we use the DC2 \texttt{truth\_match} catalogs. Entries in the \texttt{truth} catalog are matched to entries in the \texttt{object} catalog by first taking all \texttt{truth} entries within 1\arcsec of an \texttt{object} entry, and within an r-band magnitude difference of 1 mag.  Then, the \texttt{truth} entry with the smallest r-band magnitude difference is matched to a \texttt{object} entry.  If no \texttt{truth} entry meets this criterion, the closest \texttt{truth} entry is matched.   We also filter by selecting \texttt{truth\_match} entries with the flag \texttt{detect\_isPrimary}=True. This flag helps to produce a catalog of unique objects, as it filters out duplicate objects appearing in the overlapping region of multiple patches.  It also filters out isolated objects that had the deblender applied to them, in favor of using nondeblended measurements for isolated objects.

As with \textsc{DeepDISC} photo-zs, we limit the training of BPZ and FZB to objects with an i-band magnitude of less than 25.3 mag.  We use forced photometry cModel magnitudes and magnitude errors as inputs for both BPZ and FZB, and account for Milky Way dust reddening by applying a magnitude correction in each band 
\begin{equation}
    mag_{\rm dereddened} = mag - R_\lambda * E(B-V)
\end{equation} where $R_\lambda$ is calculated at the effective wavelength of each of the \textit{ugrizy} filters using the extinction law from \citep{CCM89} and E(B-V) is the reddening due to Milky Way dust.  We use the SFD dustmap available in the \texttt{dustmaps} \citep{Dustmaps} python package to obtain the E(B-V) at the location of each object. Negative flux values due to background over-subtraction and/or large noise fluctuations lead to inf or NaN values for magnitudes in a given filter. These are replaced with the 1$\sigma$ limiting magnitude of that filter at 5 years. Although this design choice is made to allow the catalog-level codes to handle negative flux measurements, we note that it biases or removes information from the catalog-level estimators and will affect photo-z likelihoods. An alternative approach worth exploring in future work would be to replace per-band negative fluxes with values sampled from a distribution derived from a normalizing flow trained on the training set \citep{PZflow}. 

\subsubsection{BPZ}

BPZ is a template-based code which uses a model to determine the likelihood of a galaxy's colors given an input set of template SEDs and a redshift grid. For BPZ photo-zs, We use a set of 200 template SEDs specialized for the simulated DC2 data.  We note that the simulated internal galactic dust extinction contained an underlying unphysical feature in its parametrization which the principle component analysis method used to construct the SED templates was not able to capture.  This feature manifests mostly at $z>1.5$, and thus high redshift performance of the BPZ algorithm will not be optimal at those redshifts, as the theoretical SEDs will not match those of the observed data.

The following notation uses magnitudes, as is the original convention from \cite{BPZ}.  However, we note that the BPZ algorithm converts input magnitudes to fluxes that are more well-behaved and have Gaussian errors. BPZ defines the posterior probability of a galaxy having redshift $z$ as $p(z|C,m_0)$ where $C$ is the galaxy colors obtained from the observed magnitudes and m$_0$ is the apparent magnitude in a single band (here, i-band). The posterior is obtained by marginalizing over a set of template SEDs
\begin{equation}
    p(z|C,m_0) \propto \sum_T p(C|z,T) p(z,T|m_0)
    \label{eq:BPZ}
\end{equation}
where $p(z|T,m_0)$ is a prior probability of a galaxy having redshift $z$ given its apparent magnitude and template type $T$ and $p(C|z,T)$ is the likelihood of a galaxy having colors $C$ given a redshift and template type.  This prior is parametrized in \cite{BPZ} and is determined from a best-fit to the data. This training of the prior, or ``informing" the algorithm, makes BPZ a competitive template-fitting code.  For more detailed functional forms of the likelihood and prior, see \cite{BPZ}. The BPZ model is informed with our input training catalog on a single CPU in 35 minutes.

\subsubsection{FZB}

FZB is a machine-learning code which projects the conditional photo-z likelihood $f(z|x)$, where $x$ represents the galaxy's observed magnitudes and magnitude errors, onto a series of orthonormal basis functions 

\begin{equation}
    f(z|x) = \sum_i \beta_i(x)\phi_i(z),
\end{equation}
in this case, a Fourier cosine basis requiring a post-hoc normalization procedure defined by algorithm-specific hyperparameters.
The expansion coefficients $\beta_i(x)$ can be determined via regression with the data using \texttt{xg-boost \citep{Chen16}.} To train the FZB model, we use a maximum of 35 cosine basis functions, and the RAIL default hyperparameters $\code{bumpmin}=0.02$, $\code{bumpmax}=0.35$, $\code{nbump}=20$, $\code{sharpmin}=0.7$, $\code{sharpmax}=2.1$, $\code{nsharp}=15$, and $\code{max\_depth}=8$ The FZB model is informed with our input training catalog on a single CPU in 24 minutes.


\section{Results}
\label{sec:results}

\begin{figure*}
    \centering
    \includegraphics[width=\textwidth]{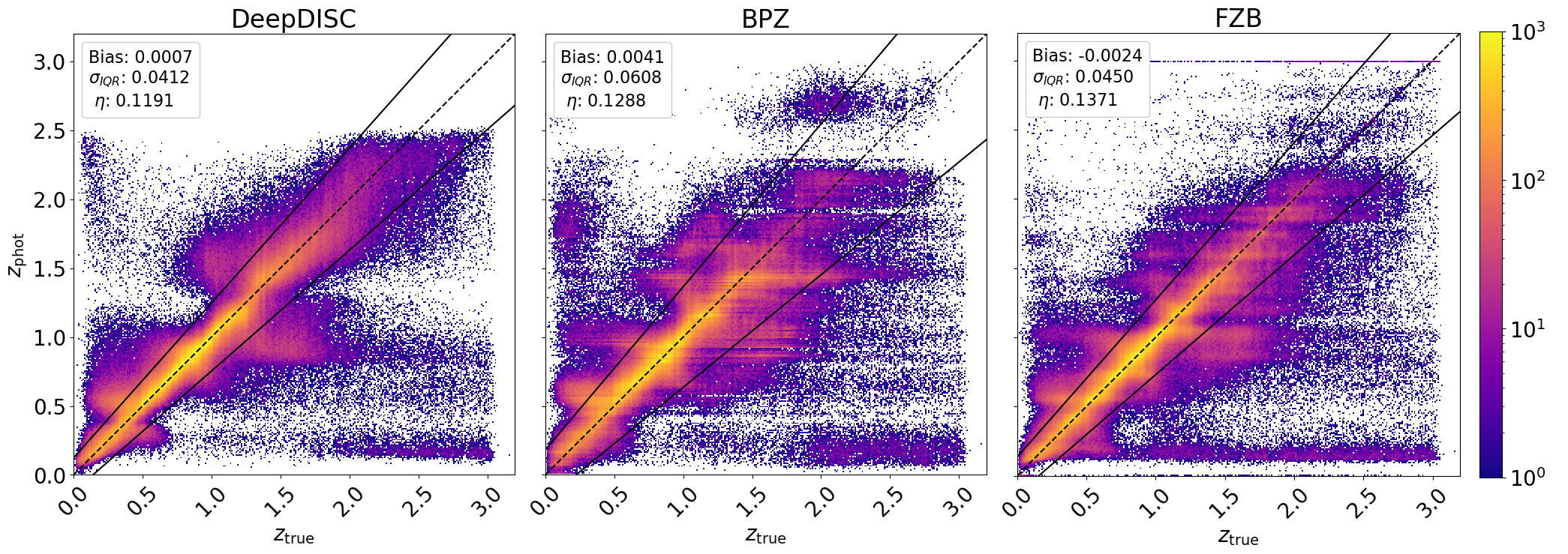}
    \caption{Redshift point estimates of \textsc{DeepDISC} (left), BPZ (middle) and \textsc{FlexZBoost} (right) compared to the true redshift. We use the mode of each photo-z PDF as our point estimate.  The dashed line is along $z_{\rm true}= z_{\rm phot}$, and the solid lines define the $3\sigma_{\rm IQR}$ outlier boundary.  Color corresponds to the number density of objects in each bin.  Bias, scatter and outlier fraction (Equations \ref{eq:bias}, \ref{eq:scatter} and  \ref{eq:outlier_frac}, respectively) are shown in the legends. }
    \label{fig:point_estimates}
\end{figure*}

\begin{figure*}
    \centering
    \subfloat{\includegraphics[width=0.5\textwidth]{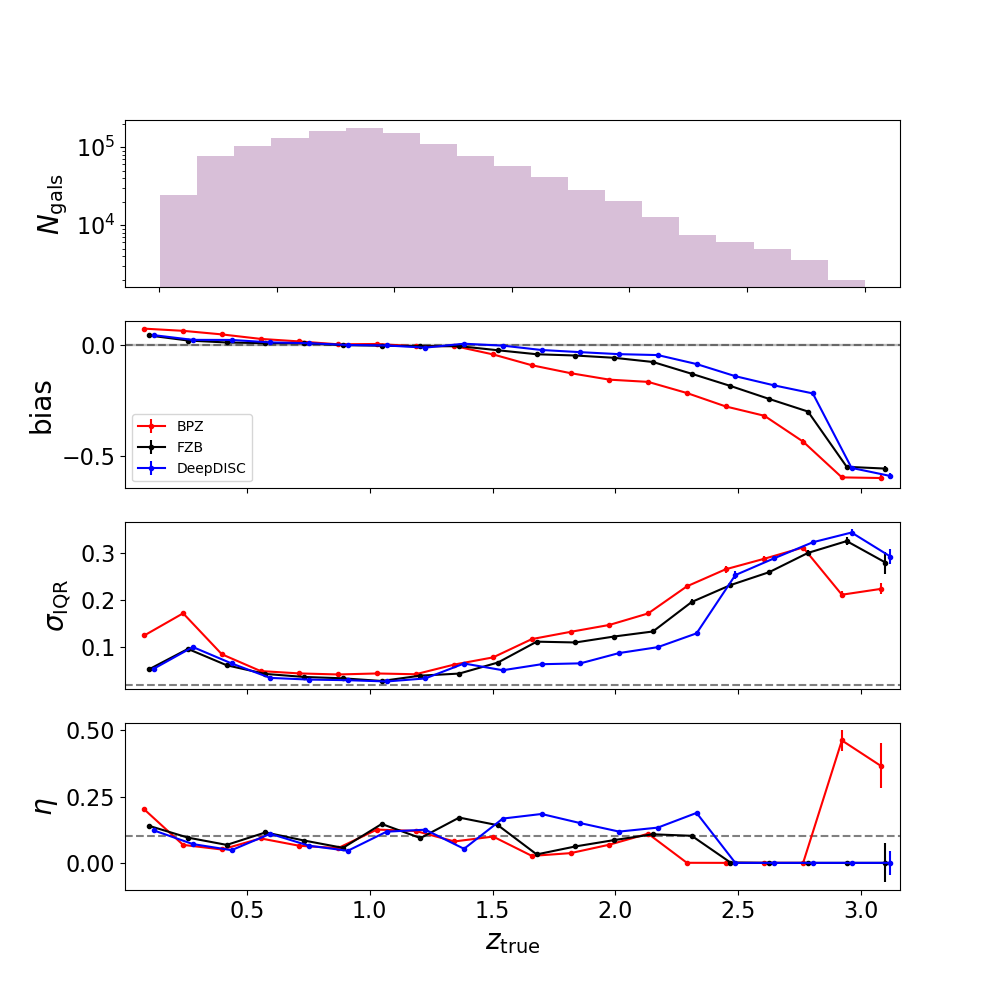}}
    \subfloat{\includegraphics[width=0.5\textwidth]{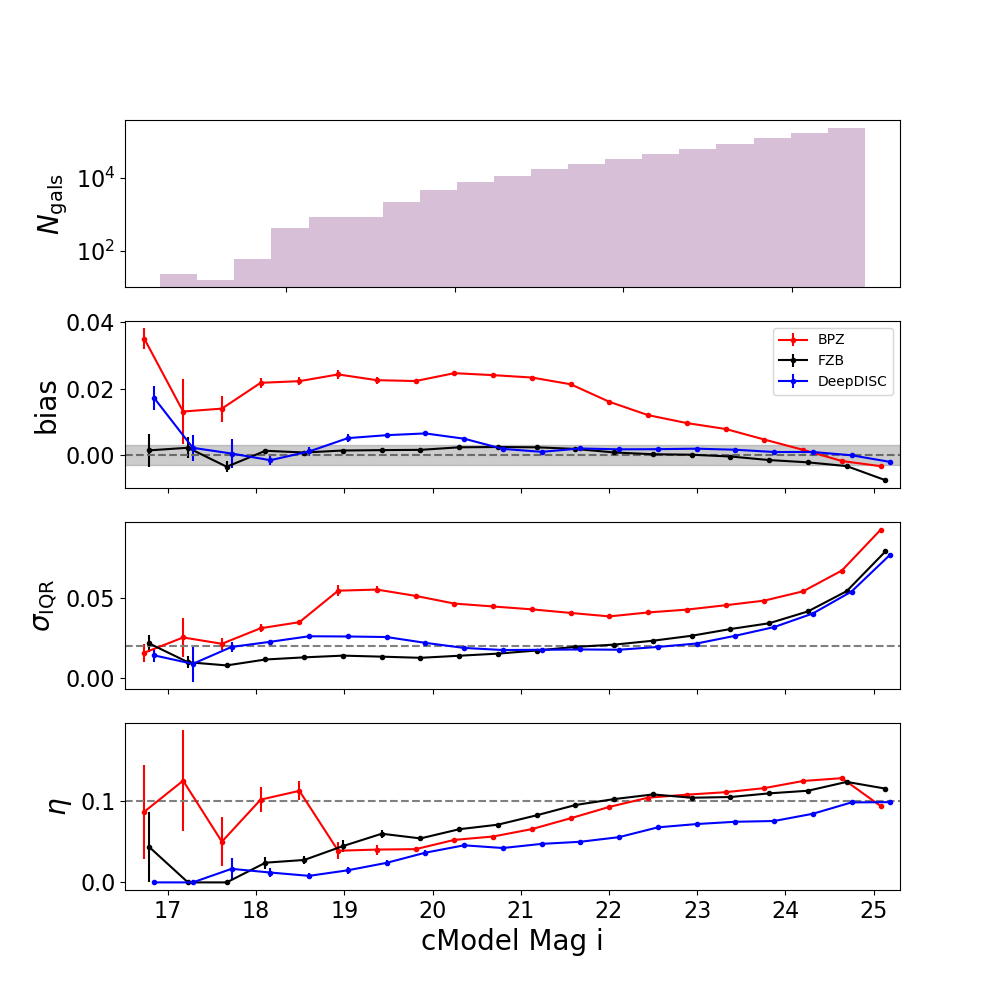}}
    \caption{Top: Histogram of galaxy redshifts in the test sample.  Lower: Point estimate metrics as a function of true redshift.   Error bars are derived from bootstrapping in each redshift bin.  The LSST science requirements are shown as a gray band in the second panel and grey dashed lines in the lower two panels.  Bias ideally falls within the gray band, whereas $\sigma_{\rm IQR}$ and $\eta$ ideally lie below the gray lines.} 
    \label{fig:sys-ztrue}
\end{figure*}

In this section, we compare all codes and quantify the quality of our photo-z estimation results using several  metrics designed to measure the accuracy of point estimates derived from the PDFs.  After training \textsc{DeepDISC}, we apply the model to our test set of images. This takes about 40 minutes to run on a single NVIDIA V100 GPU and produces a catalog of 3,177,538 objects. We compute object locations using the centers of predicted bounding boxes, and use each image's world coordinate system (WCS) to convert these pixel values to RA and DEC coordinates. In order to compare to the catalog-based codes, we cross-match to the DC2 \texttt{object} catalog by taking the closest gold sample match within 1\arcsec. This produces a test set of 1,254,379 objects for comparison.  This detection catalog is 94.2\% complete with gold sample objects in the DC2 \texttt{truth} catalog and 96.9\% complete with the gold sample of the DC2 \texttt{object} catalog.  We find that \textsc{DeepDISC} detection completeness tends to decrease with brighter i-band magnitudes, which may be due to both overdeblending, or large, bright objects being truncated by the image cutout.

We then obtain BPZ and FZB PDFs by applying these models to the cModel magnitudes (corrected for Milky Way dust reddening) and errors of the matched test set.  As with the training set, the $1\sigma$ limiting magnitude is used for instances of non-detections or pipeline failures.  We parallelize BPZ and FZB estimation over 16 CPUs.  BPZ takes 31.64 minutes and FZB takes 5.08 minutes to estimate the photo-z PDFs of all test set objects.   

We use the mode of the PDF as our point estimate $z_{\textrm {phot}}$, and quantify the bias $e_z$ in our estimates by
\begin{equation}
    \label{eq:bias}
    e_z = (z_{\textrm{phot}} - z_{\textrm{true}}) / (1+ z_{\textrm{true}}).
\end{equation}
We also quantify the scatter of the estimates as $\sigma_{\textrm{IQR}}$.  This is the interquartile range, or 
\begin{equation}
    \label{eq:scatter}
    \sigma_z = (e_{z75} - e_{z25}) / 1.349 
\end{equation}
where $e_{z75}$ and $e_{z25}$ are the 75th and 25th percentile of $e_z$, and we divide by 1.349 to ensure that the area spanned by $\sigma_{\rm IQR}$ is equivalent to the area within one standard deviation for a standard Normal distribution. The outlier fraction is given by
\begin{equation}
    \label{eq:outlier_frac}
    \eta = N_{\textrm{out}}/N_{\textrm{tot}}
\end{equation}
with N$_{\textrm{out}}$ as the number of galaxies with $|e_z| > max(3\sigma_{\textrm{IQR}}, 0.06)$, following the definition of \citeDCt.

A 2D histogram of $z_{\textrm{true}}$ vs $z_{\textrm{phot}}$ for all three codes is shown in Figure \ref{fig:point_estimates}. In general, the dense regions lie along the $z_{\textrm{true}}$ vs $z_{\textrm{phot}}$ line, with catastrophic outliers filling in other regions of the plot.  The bias, scatter, and outlier fraction for the entire sample are listed in Table \ref{tab:point_metrics}.  Overall, \textsc{DeepDISC} performs better than BPZ or FZB.  The performance as a function of true redshift is shown in Figure \ref{fig:sys-ztrue}, with error bars derived from the standard deviation of a bootstrapped distribution of $z_{\rm phot}$ per $z_{\rm true}$ bin.  Each code has distinct failure modes, highlighting the challenge of photometric redshift estimation.  

At low redshifts, \textsc{DeepDISC} and FZB perform the best in terms of bias and scatter.  Despite this region being underrepresented in the training as there are relatively few low-z objects in the training sample N(z) (see Figure \ref{fig:trainz_hist}), \textsc{DeepDISC} is able to perform on par with or better than \textsc{FlexZBoost}. BPZ performs the worst at this low redshift range.  This is likely due to a degeneracy between the Balmer break at low-z and Lyman break at high-z causing confusion between high and low-z objects, as seen in Figure \ref{fig:point_estimates}.  

At mid-range redshifts, the codes show similar performance, with \textsc{DeepDISC} maintaining a slightly lower bias and scatter.  All codes exhibit a trend towards larger negative bias and larger scatter as redshift increases.  This is expected, as high-z objects are underrepesented in the training and are susceptible to the fact that the Balmer break transitions out of the y-band at z$\sim$1.4. As mentioned in \ref{subsec:cat-codes}, the BPZ template SEDs are not completely representative at high-z due to unphysical dust extinction, which likely contributes to its degraded performance seen at these redshifts. At $1.5\leq z \geq2.5$, \textsc{DeepDISC} maintains a significantly lower scatter and absolute bias value, indicating that the high dimensional space of features it learns from the images encodes more information than the photometry.  We discuss this point more in Section \ref{subsec:obs-sys} and Appendix \ref{app:blurred}.

BPZ and FZB see a decrease in scatter beyond z$\sim$2.5, at which the Lyman break enters the u-band. However, all codes underpredict the redshifts at high-z.  Some poor performance at the edges of the redshift distribution is expected, due to noise and the choice of redshift grid used to parametrize the PDFs.  Additionally, mode point estimates fail in the case of highly bimodal PDFs (high redshift blue galaxies may be confused for low redshift red galaxies).  However, \textsc{DeepDISC} is especially sensitive to this regime, as no photo-z PDFs produced by \textsc{DeepDISC} yield a mode above z$\sim$2.5.  During training, there is on average less than one gold sample object with z>2.5 per image.  Many different attempts to address this issue were made, including weighting the redshift loss function in proportion to the redshift distribution N(z), resampling predicted regions produced by the Region Proposal Network to a more uniform redshift distribution, and doubling the training set size and number of epochs. It appears that the information that \textsc{DeepDISC} is able to extract from the images of sources is more meaningful for photo-z estimation when compared to source photometry. However, it is susceptible to the number of objects at a given redshift in the training set.  We discuss this result and possible solutions more in Section \ref{sec:discussion}.

Point estimate metrics as a function of i-band cModel magnitude are shown in the right panel of Figure \ref{fig:point_estimates}.  In general, scatter and outlier fraction increase with magnitude for all codes.  This is expected, as the signal-to-noise for faint objects is small.  \textsc{DeepDISC} and FZB tend to perform better than BPZ at bright magnitudes despite the smaller sample size of objects. The larger bias of BPZ photo-zs at bright magnitudes reflect a phenomenon with template fitting codes due to the smaller photometric uncertainties measured at bright magnitudes.  Smaller errors can lead to larger $\chi^2$ values in the likelihood fitting process which exaggerate the importance of biased flux values. This will be especially noticeable if the photometric errors measured by the LSST science pipeline are underestimated. 

\begin{figure*}
    \centering
    \subfloat{\includegraphics[width=0.3\textwidth,height=7cm]{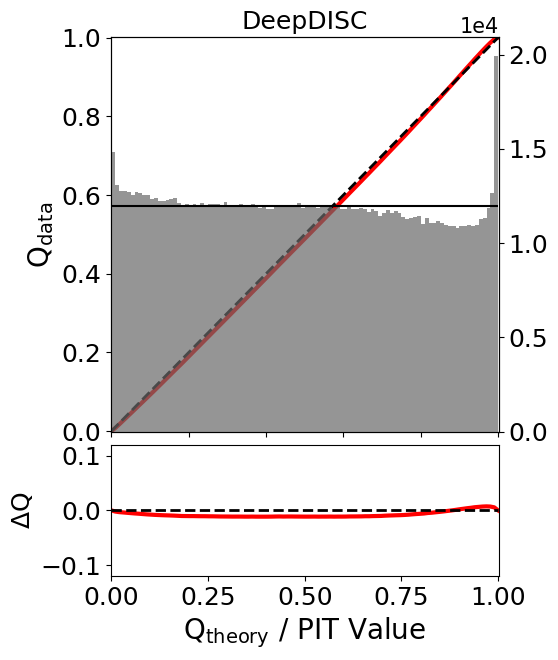}}
    \subfloat{\includegraphics[width=0.3\textwidth,height=7cm]{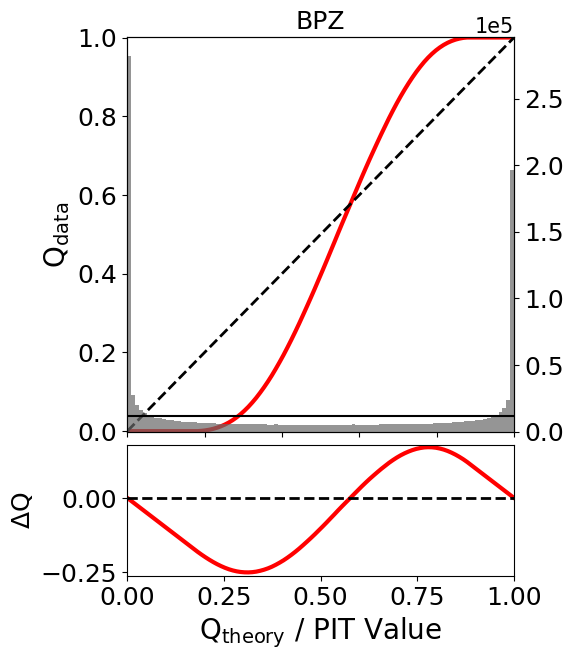}}
    \subfloat{\includegraphics[width=0.3\textwidth,height=7cm]{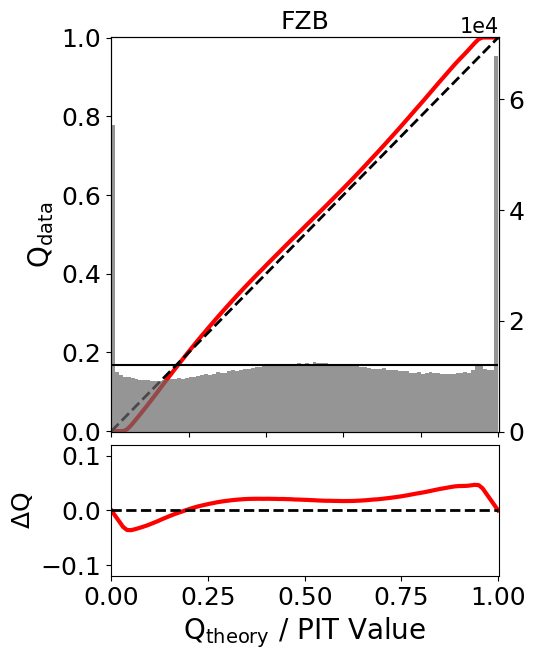}}
    \caption{PIT histograms and QQ plots of the PDF ensembles produced by each code on the test set. PIT histograms are shown in grey and correspond to the right y-axis.  The black horizontal line represents a uniform PIT histogram.  The red line plots quantiles of the ensemble PIT values compared to quantiles of an ideal flat distribution and corresponds to the left y-axis. The black dashed line represents $\rm Q_{\rm data}=\rm Q_{\rm theory}$.  Lower panels show the a zoomed in difference between $\rm Q_{\rm data}$ and $\rm Q_{\rm theory}$. DeepDISC produces the most visually uniform PIT distribution.}
    \label{fig:PIT_hists}

\end{figure*}

\begin{figure}
    \centering
    \includegraphics[width=0.9\columnwidth]{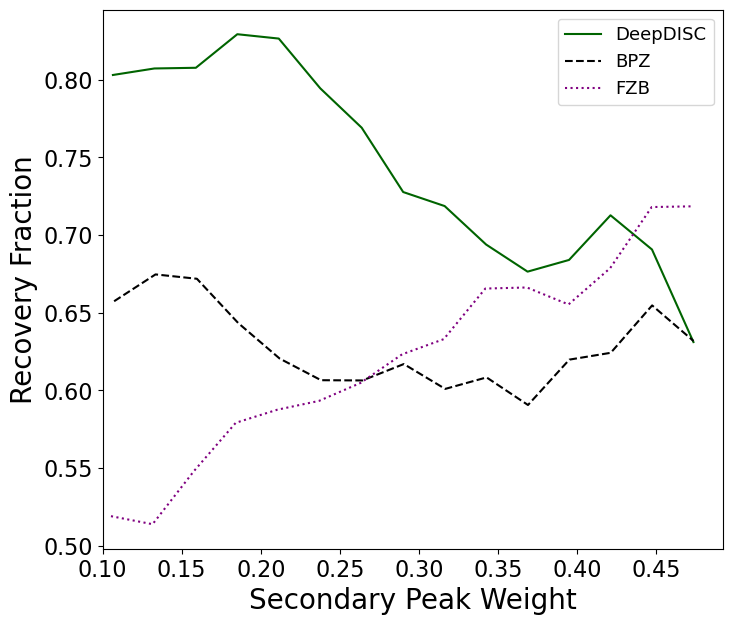}
    \caption{Recovery fraction, i.e., fraction of outliers that are no longer considered outliers if their PDF is evaluated at the secondary peak rather than the primary.  We only include secondary peaks with a weight of at least 0.1 in the Figure, and plot the peak weight on the x-axis.  The high recovery fractions indicate that the PDFs are encoding the true redshift even if the primary peak is not a reliable point estimate.}
    \label{fig:rf}
\end{figure}

\begin{table*}
    \centering
    \begin{tabular}{cccc|ccccc}
        \hline
        \hline
         & bias & $\sigma_{\textrm{IQR}}$ & $\eta$  & cde loss & KS & CvM & AD  \\ 
         \hline
         \textsc{DeepDISC} &  0.0007 & 0.0412 & 0.1191 & -4.249 & 0.0115 & 100.6 & 455.3 \\
         BPZ &  0.0041 & 0.0608 & 0.1288 & 0.5973 & 0.1455 & 9061 & 50260 \\
         FZB &  -0.0024 & 0.0450 & 0.1371 & -3.577 & 0.0430 & 498.7 & 6474 \\
         \hline
    \end{tabular}
    \centering
    \caption{Point estimate and PDF metrics on the test set pz estimates for each code.  \textsc{DeepDISC} outperforms BPZ and FZB in all cases.}
    \label{tab:point_metrics}
\end{table*}


%

\subsection{PDF metrics}
We use probability integral transform (PIT) histograms to visually judge the calibration of our PDFs, i.e., if the errors our model produces are reliable.  The PIT histogram is a visual diagnostic tool to determine whether the PDFs in an ensemble are too broad or too narrow (\citet{DIsanto18}, \citeDCt).  The PIT value for each PDF is calculated as the cumulative distribution function (CDF) evaluated at the true redshift.
\begin{equation}
    PIT \equiv CDF(z_{\textrm{true}}) = \int_{-\infty}^{z_{\textrm{true}}} PDF(z) dz
\end{equation}
An ideal PIT histogram is a uniform distribution, as true redshifts should be randomly sampled from their underlying distributions and thus give uniform CDFs.  Overly narrow (overconstrained) PDFs will tend to produce a PIT histogram with an abundance of values at 0 and 1, as there is no probability mass in the region of the true redshift.  Overly broad (underconstrained) PDFs tend to contain near equal amounts of probability for redshifts $z<z_{\textrm{true}}$ and $z>z_{\textrm{true}}$, and thus produce a PIT histogram with a bulge around 0.5.  The ideal PIT histogram follows a uniform distribution, indicating neither too broad nor too narrow uncertainties.  Systematic biases can be seen as deviations from uniformity. The quality of the PDFs can be seen in Figure \ref{fig:PIT_hists}.  \textsc{DeepDISC} produces the least outliers at PIT values close to 0 and 1, and appears to produce the most uniform PIT histogram.  The BPZ PIT histogram is noticeably overconstrained, indicating BPZ tends to underestimate the uncertainty on its redshift estimates.  This may also be due to underestimated photometric uncertainties which are measured from the simulated coadded images using the LSST Science pipelines software.

As a further test of the PDF quality, we calculate the CDE loss, Kolmogorov-Smirnov (KS), Cramer von Mises (CvM), and Anderson-Darling (AD) statistics for each code.  The CDE loss is estimated (up to a constant) by 
\begin{equation}
    CDE = \mathbb{E}\left[\int{f(z|x)^2}dz \right] - 2\mathbb{E}_{Z,X}\left[f(Z|x)\right]
\end{equation}
where $f(z|x)$ is the estimated photo-z PDF given the input observables, $Z$ is the space of all possible redshifts (in practice the grid of values on which the PDF estimates are binned), and $\mathbb{E}$ is the expectation value.  This metric is analogous to a root-mean-square error in traditional regression problems, and a large, negative CDE loss is optimal.  The KS, CvM and AD statistics all quantify how close an empirical distribution of CDFs $\hat{y}$ is to the ideal theoretical distribution of CDFs, which is in our case a uniform distribution $y = \mathcal{U}(0,1)$.  The KS statistic is given as 
\begin{equation}
    KS = max\big(|CDF[\hat{y},z] - CDF[y,z]|\big)
\end{equation}
where $\hat{y}$ is the empirical distribution of PIT values. The CvM statistic is
\begin{equation}
    CvM^2 = \int_{-\infty}^{\infty} (CDF[\hat{y},z] - CDF[y,z])^2 dCDF[y,z]
\end{equation}
and since it is a sum of squared differences, is more sensitive to large deviations and many deviations over the entire empirical CDF distribution. 
Lastly, the AD statistic
\begin{equation}
    AD^2 = N_{tot} \int_{-\infty}^{\infty} \frac{(CDF[\hat{y},z] - CDF[y,z])^2}{(CDF[\hat{y},z](1-CDF[y,z])}dCDF[y,z],
\end{equation}
which is more sensitive to the tails of the empirical CDF distribution. Since all of these PDF metrics quantify deviations from the ideal theoretical distribution, lower is better. We list the these metrics for each code in Table \ref{tab:point_metrics}.  \textsc{DeepDISC} produces the lowest (best) value of each of these metrics for the test set.  

To further investigate the reliability of \textsc{DeepDISC} PDF estimates, we examine the multimodal nature of the PDFs encoded by the MDN. We use a peak finding algorithm to determine secondary modes of the PDFs, and examine whether the secondary modes capture meaningful information.  We find that 23.5\% of the \textsc{DeepDISC} PDFs have secondary peaks.  We define the relative peak weight for each PDF $p(z)$ by $\frac{p(z_2)}{\sum_{i} p(z_i)}$ where $z_i$ are the peaks.  13.3\% of \textsc{DeepDISC} PDFs have a secondary peak with a weight above 0.1.  For BPZ, 61.5\% of the PDFs have a secondary peak and 37.9\% of the PDFs have a weight above 0.1.  FZB yields 52.4\% of PDFs with secondary peaks and 43.7\% with the peak weight above 0.1.  Given the relatively simple nature of the \textsc{DeepDISC} Gaussian mixture model parametrization, a lower number of secondary peaks compared to the other codes is not unexpected.  However, to see whether these secondary modes contain meaningful information, we evaluate the recovery fraction for these PDFs.  The recovery fraction is defined as the fraction of photo-z outliers that no longer qualify as outliers if $z_{\textrm{phot}}$ is taken to be at the secondary mode of the PDF rather than the first mode.  We show the recovery fraction as a function of secondary peak weights in Figure \ref{fig:rf}, limiting the analysis to PDFs with secondary peaks with a weight above 0.1. For all codes, recovery fractions are above 50\% for strong and weak peaks, indicating that the photo-z estimators are able to capture meaningful degeneracies in the PDFs.

Overall, we have shown that under conditions of closely matched prior information, \textsc{} is very competitive in its photo-z estimation.  In our experimental setup, \textsc{DeepDISC} photo-z outperforms traditional photo-z codes in terms of bias, scatter, and outlier fraction.  Going beyond single point metrics, we examine the quality of the errors encoded in the PDFs by looking at the PIT histograms and additional metrics designed to measure how well-calibrated our ensemble of PDFs is. We again find that \textsc{DeepDISC} produces a PIT distribution closest to the ideal uniform, indicating that  errors are neither too broad nor too narrow. We also find that for all codes, a majority of outliers have coverage at the secondary peak of their PDFs, suggesting that the models captures physically meaningful degeneracies.  In the next Sections, we examine possible limiting factors of our model and dependencies of our results.  We examine the effects of imaging systematics, the dependence on data quality, and possible scaling laws in the context of our model. 

\subsection{Observational Systematics}
\label{subsec:obs-sys}

\begin{figure*}[htbp]
    \centering
    \includegraphics[width=0.55\textwidth]{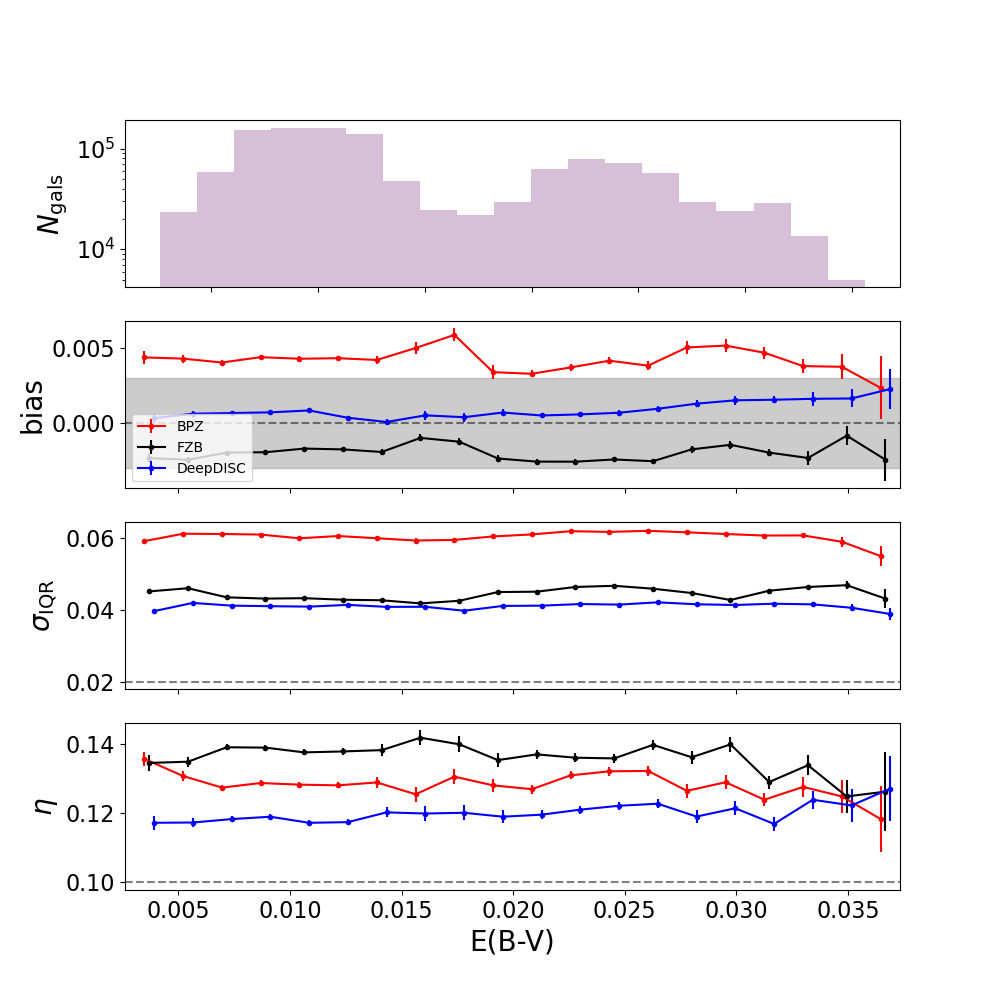}
    \caption{Top: Histogram of Milky Way dust reddening E(B-V) for objects in the test sample. Lower: Bias, scatter, and outlier fraction metrics as a function of E(B-V). Error bars are derived from bootstrapping in each bin.  The LSST science requirements are shown as a gray band in the second panel and grey dashed lines in the lower two panels.  A slight positive correlation with bias appears in \textsc{DeepDISC} estimates, although it remains below the LSST statistical error budget. Outlier fractions also slightly increase with E(B-V). }
    \label{fig:ebv_comp}
\end{figure*}

\begin{figure*}[htbp]
    \centering
    \includegraphics[width=0.55\textwidth]{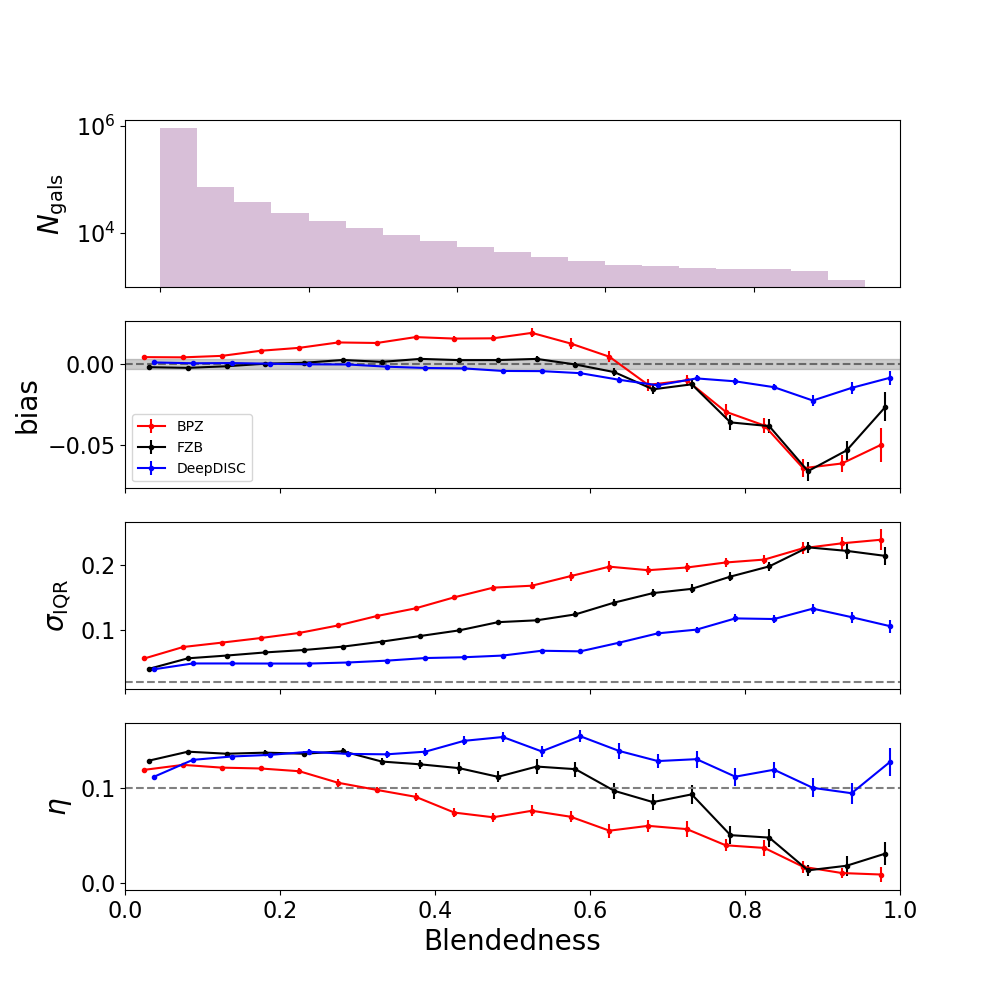}
    \caption{Top: Histogram of blendedness of objects in the test sample. Lower: Bias, scatter, and outlier fraction metrics as a function of blendedness. Error bars are derived from bootstrapping in each bin.  The LSST science requirements are shown as a gray band in the second panel and grey dashed lines in the lower two panels.   The bias and scatter of BPZ and FZB photo-z estimates significantly worsen as blending increases, whereas DeepDISC photo-z bias and scatter are much less sensitive to blending. }
    \label{fig:blendedness_comp}
\end{figure*}

\begin{figure*}[htbp]
    \centering
    \includegraphics[width=0.55\textwidth]{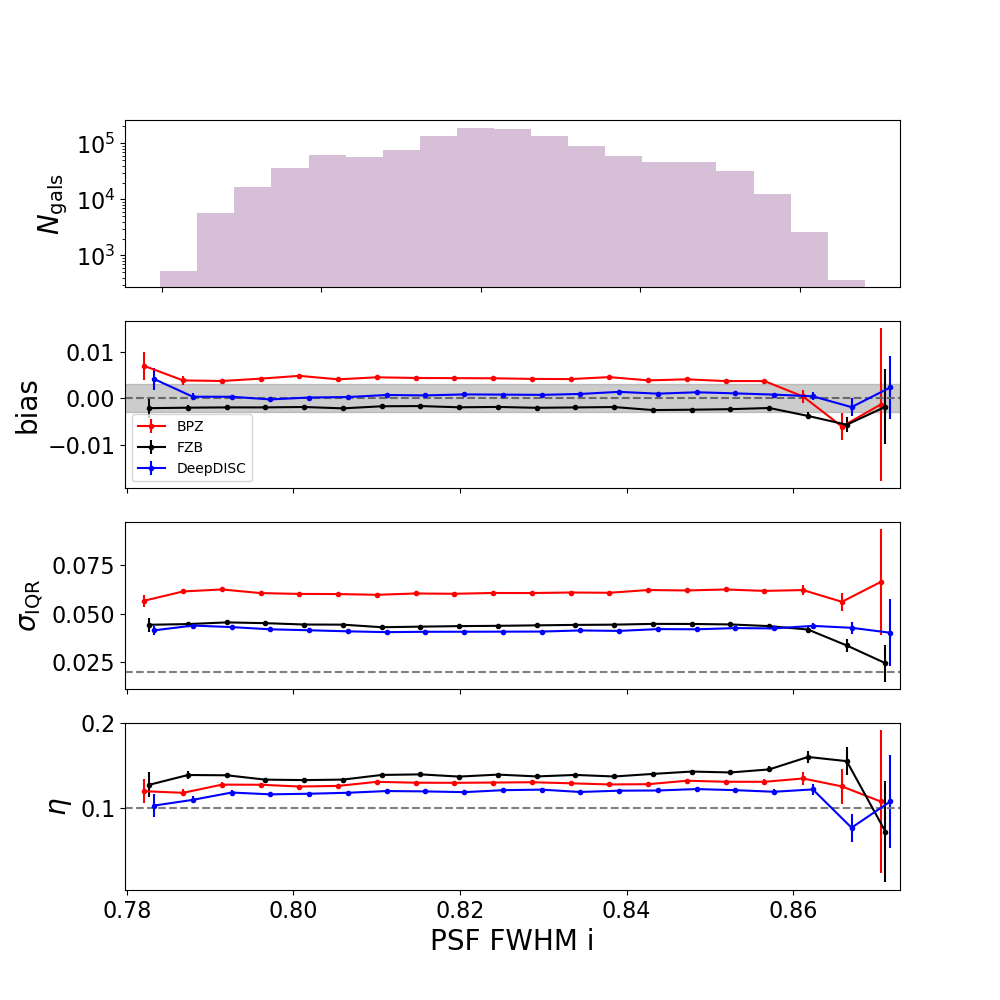}
    \caption{Top: Histogram of PSF FWHM in the i-band for objects in the test sample. Lower: Bias, scatter, and outlier fraction metrics as a function of PSF FWHM. Error bars are derived from bootstrapping in each bin.  The LSST science requirements are shown as a gray band in the second panel and grey dashed lines in the lower two panels.   All models remain largely insensitive to PSF FWHM, although BPZ's scatter shows a slightly positive trend. }
    \label{fig:psfi_comp}
\end{figure*}

Due to the many exposures that will be collected by LSST over its observing run, the image coaddition process will result in an averaging over several systematics.  However, it is still important to characterize the performance of our photo-z estimator as a function of imaging systematics to understand weaknesses and biases that may propagate to downstream measurements.  In this work, we have only explicitly accounted for Milky Way dust reddening by adding the E(B-V) values as extra input to the redshift ROI head of \textsc{DeepDISC}.  We examine photo-z performance as a function of E(B-V) as well as point spread function (PSF) variations and blending in Figures \ref{fig:ebv_comp}, \ref{fig:blendedness_comp} and \ref{fig:psfi_comp}.

Milky Way dust in the line of sight of a source will scatter and absorb light and thermally re-emit in the infrared.  Observationally, this will appear as a dimming and reddening \citep{Draine03}. For catalog-based photo-z codes that use measured fluxes or magnitudes, it is important to correct for this effect by using the measured reddening E(B-V) and extinction law for each photometric filter.  The DC2 area was chosen to be at high Galactic latitude in a relatively uniform, low-dust region, which results in the small maximum E(B-V) and apparent bimodality in the distribution in Figure \ref{fig:ebv_comp}. We see a slightly positive trend in bias vs E(B-V) for \textsc{DeepDISC} and an uptick in outlier fraction at the highest E(B-V) values.  However, the bias remains minimal and below the LSST science requirements error budget.  No strong trend is seen in scatter.  FZB and BPZ also appear largely insensitive to E(B-V), which is expected due to the aforementioned dust correction.

In order to deblend sources, the LSST science pipeline determines footprints, or image regions of above-threshold flux that contain at least one peak.  Multiple peaks in a footprint are dubbed "child" sources.  The deblending algorithm attempts to extract the light profile associated with each peak in a footprint. Blendedness is a measure of how much the flux of a deblended object may be affected by neighboring objects in the footprint  
\begin{equation}
    blendedness = 1 - \frac{flux_{\rm child}}{flux_{\rm parent}}
\end{equation}
where $flux_{\rm child}$ is the deblended child flux and $flux_{\rm parent}$ is the total flux in the footprint.  Isolated sources thus have a blendedness of 0, and a blendedness above 0.2 can be considered a significant blend.  For details of the blendedness metric, see \cite{Bosch18}.
The effect of blending on all codes is shown in Figure \ref{fig:blendedness_comp}. Up to a blendedness of around 0.5, BPZ bias increases, FZB bias slightly increases, and DeepDISC bias slightly decreases.  However, beyond a blendedness of 0.5, BPZ and FZB bias start to dramatically decrease, while DeepDISC bias only slightly decreases.  For all codes, photo-z scatter increases as a function of blendedness, although DeepDISC is the least sensitive to this systematic.  It appears that the features extracted by DeepDISC are more robust to blending than the deblended cModel magnitudes produced by the LSST science pipelines deblender.  This may be due to the fact that all pixel-level information is considered in DeepDISC feature extraction, which provides more information regarding source light profiles and SEDs than aperture magnitude measurements.  We discuss this more in Section \ref{sec:discussion} and Appendix \ref{app:blurred}.  Additionally, raising the positive IOU threshold discussed in Section \ref{sec:methods} could prevent more detection/truth mismatches during training, which could further improve the robustness to this systematic.

We examine any systematics trends due to i-band PSF FWHM in Figure \ref{fig:psfi_comp}.  The PSF characterizes the distortion of an observed image due to atmospheric effects such as refraction.  Traditional image analysis pipelines often explicitly use the PSF as input, potentially giving them an advantage over neural network approaches that do not, as is the case with \textsc{DeepDISC.}  Since \textsc{DeepDISC} uses image data and thus is encoding pixel-level features and distortions, we examine if there is any dependence in performance with PSF FWHM. No clear trends can be seen in Figure \ref{fig:psfi_comp}, except a potential dip in outlier fraction at high values. This is encouraging, in that it shows that training with a variety of images and using a ``data-driven" approach is sufficient to account for (small) PSF variations rather than explicitly inputting this information into the network. 

Overall, \textsc{DeepDISC} shows little sensitivity to these systematics, comparable to the catalog-based methods which incorporate systematics mitigation during pre-processing, e.g., deblending, or post-processing, e.g., correcting for reddening.  We find that \textsc{DeepDISC} in particular is more robust to blending than the catalog-based codes, with significantly less photo-z scatter at high levels of blending.  In future applications with real data, a variety of images taken from a wide footprint should be used in order to help the network marginalize over varying systematics. Understanding the photo-z uncertainties due to imaging systematics will be an important endeavor for downstream science cases.

\subsection{Dependence on Image Depth}
\label{sec:depth}

\begin{figure*}[htbp]
    \centering
    \includegraphics[width=0.8\textwidth]{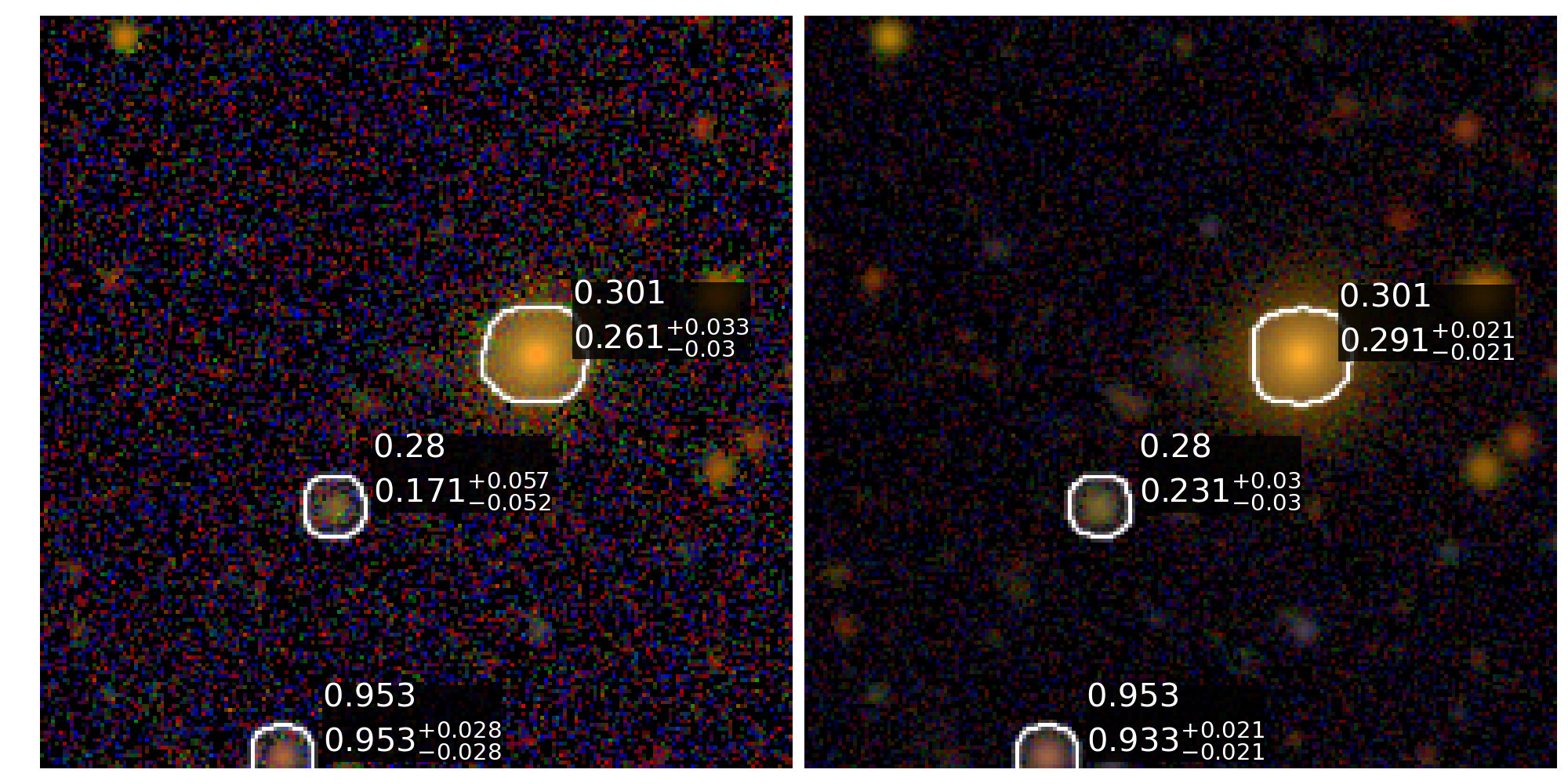}
    \caption{DeepDISC photo-z estimates on objects in a Year 1 (left) and Year 5 (right) image.  RGB colors correspond to the i,r, and g-band, respectively.  A Lupton asinh scaling is used for visualization. A subset of randomly selected objects in the images is shown.  The labels present the true redshift of the object on top, and list below the mode of the predicted PDF and errors given by the range of the 25th and 75th quantiles from the median. Overall, increased observing time leads to better photo-zs and smaller errors on the estimates. }
    \label{fig:Y1_Y5_comp}
\end{figure*}

\begin{figure*}[htbp]
    \centering
    \includegraphics[width=0.8\textwidth]{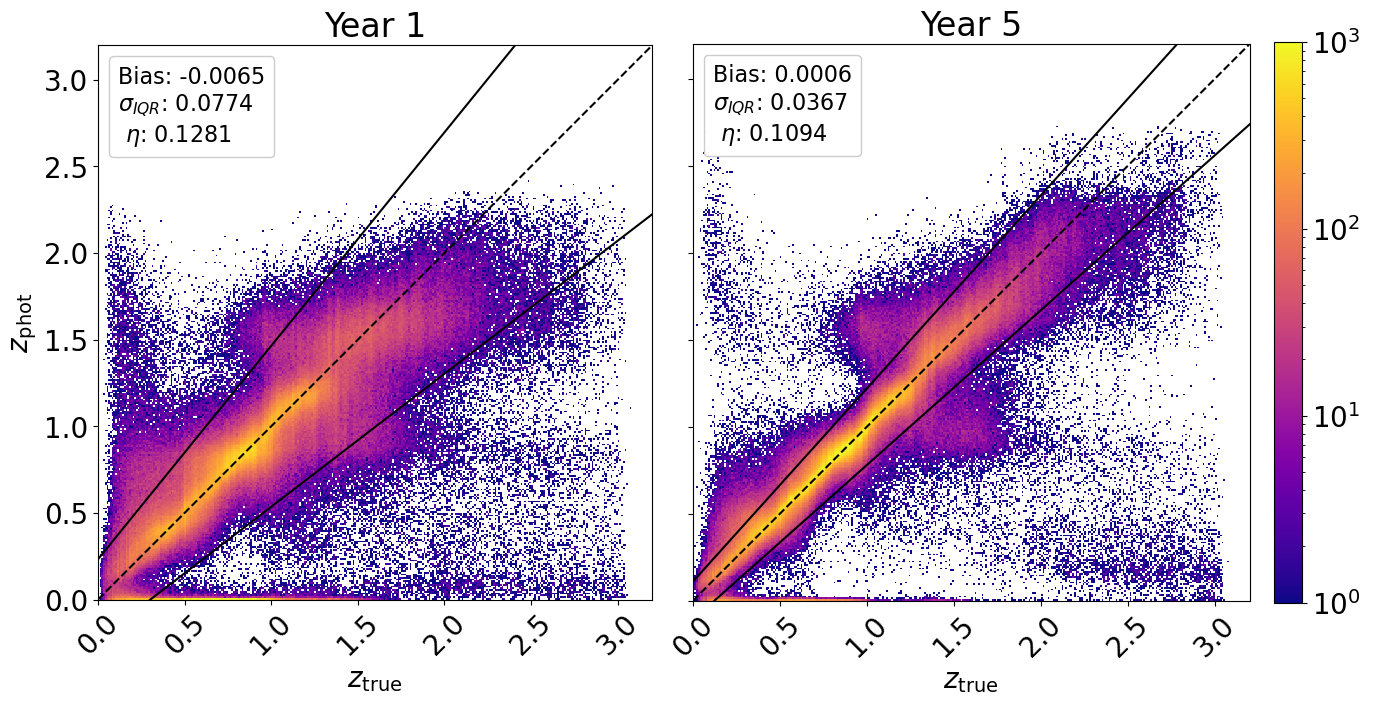}
    \caption{Redshift point estimates for Year 1 (left) and Year 5 (right) data.  The mode is used to produce a single point estimate for each pdf.  The dashed line is along $z_{\rm true}= z_{\rm phot}$, and the solid lines define the $3\sigma_{\rm IQR}$ outlier boundary.  Color corresponds to the number density of objects in each bin.  Bias, scatter and outlier fraction (Equations \ref{eq:bias}, \ref{eq:scatter} and  \ref{eq:outlier_frac}, respectively) are shown in the legends. The comparison is made on the common set of objects detected by both models. Year 5 data greatly improves the estimates overall, and notably improves the high-z regime.}
    \label{fig:depth_point_estimates}
\end{figure*}

Simulated DC2 images are available for both 1 year and 5 years of observations (Figure \ref{fig:Y1_Y5_comp}) . We compare the results of using the 1 year vs 5 year observations to see how \textsc{DeepDISC} photo-z results scale with depth in Figure \ref{fig:depth_point_estimates} and Table \ref{tab:depth_point_metrics}.  The only difference in the runs is the input imaging data.  After training and inference, we take the intersection of the sets of objects detected by each model.  This is to ensure that we are ignoring any effects from evaluating on different samples of objects, and judging solely the effect of increased signal in the images.  We find that after 5 years of observations, the bias changes from -0.0039 to -0.0001.  The scatter changes from 0.0756 to 0.0363, a factor of almost one half, and the outlier fraction evolves from 0.1161 to 0.0967.  Given that the signal-to-noise ratio (SNR) is proportional to the square root of the observing time, we see that the scatter in photo-z mode predictions is roughly inversely proportional to the SNR of the data.  Similarly, the CDE loss also scales proportionally to the SNR.  Notably, \textsc{DeepDISC} can produce more high redshift photo-z estimates when trained with the 5 year data, as seen in Figure \ref{fig:depth_point_estimates}. 
\begin{table}[htbp]
    \centering
    \begin{tabular}{cccc|ccc}
        \hline
        \hline
         & bias & $\sigma_{\rm IQR}$ & $\eta$  & cde loss \\ 
         \hline
         \textsc{DeepDISC} Y1 & -0.0039  & 0.0756 &  0.1161 & -2.280 \\
         \textsc{DeepDISC} Y5 & -0.0001  & 0.0363 &  0.0967 & -4.674 \\
         \hline
    \end{tabular}
    \centering
    \caption{Comparing \textsc{DeepDISC} photo-zs when using the 1 year vs 5 year observations.  Same metrics as Table \ref{tab:point_metrics}.}
    \label{tab:depth_point_metrics}
\end{table}

\subsection{Model Scalability}
\label{subsec:scaling}
Scaling laws describe model performance as a function of increased training set size or model size.  They are a useful tool for experimental design, especially in the case of limited training data or compute resources.  It is important to be able to quantify potential gains and evaluate trade-offs as models increase in size and compute resource requirements. Generally, it has been found that increasing the model size and/or training set size improves performance of neural networks \citep{Zhai21,Hestness17}.  However, these findings typically stem from experiments with ImageNet or other terrestrial datasets.  Investigation of scaling laws into astronomical data sets is fairly under-explored \citep{Walmsley24, Smith24}.  

\begin{figure}[htbp]
    \centering
     \subfloat{\includegraphics[width=0.35\textwidth]{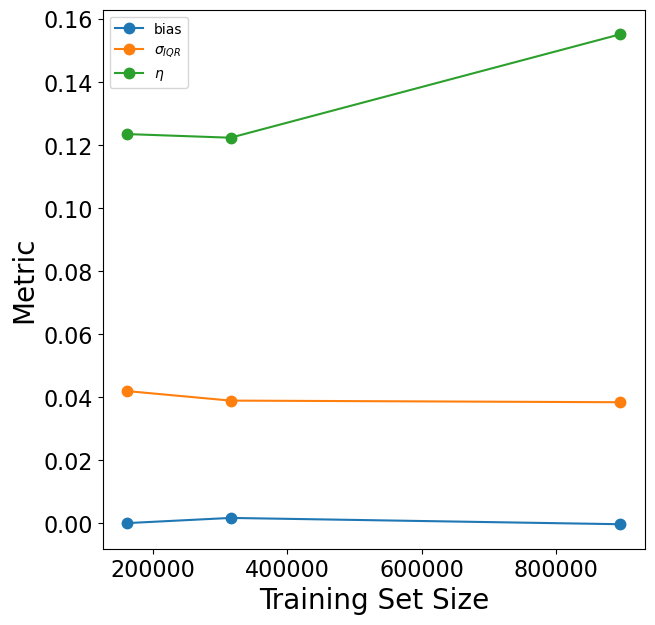}} \quad
     \subfloat{\includegraphics[width=0.35\textwidth]{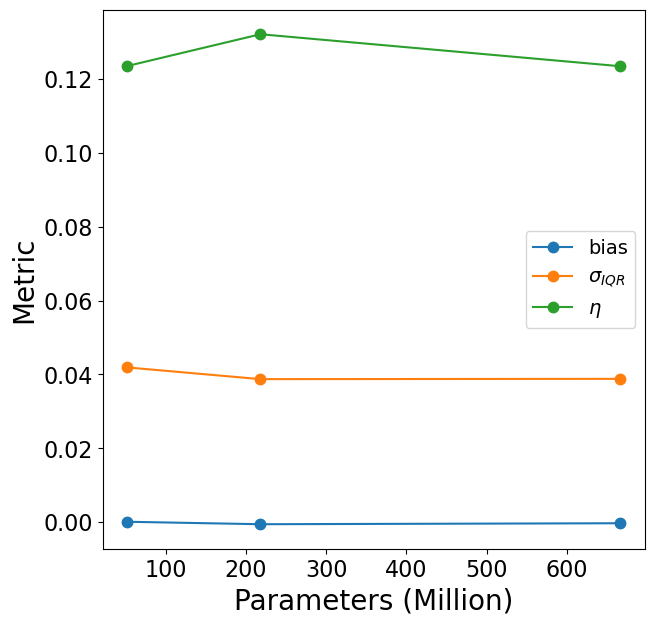}}
    \caption{Scaling laws for redshift point estimates.  There appears to be no obvious relationship with model or training set size and point estimate metrics. The top panel shows the effect of increasing the training set size by 2x and 5x, and the bottom panel shows the effect of using larger model backbones, in our case the MViTv2 Base, Large, and Huge models.}
    \label{fig:scaling_laws_model}
\end{figure}

\begin{figure*}
    \centering
    \subfloat{\includegraphics[width=0.45\textwidth]{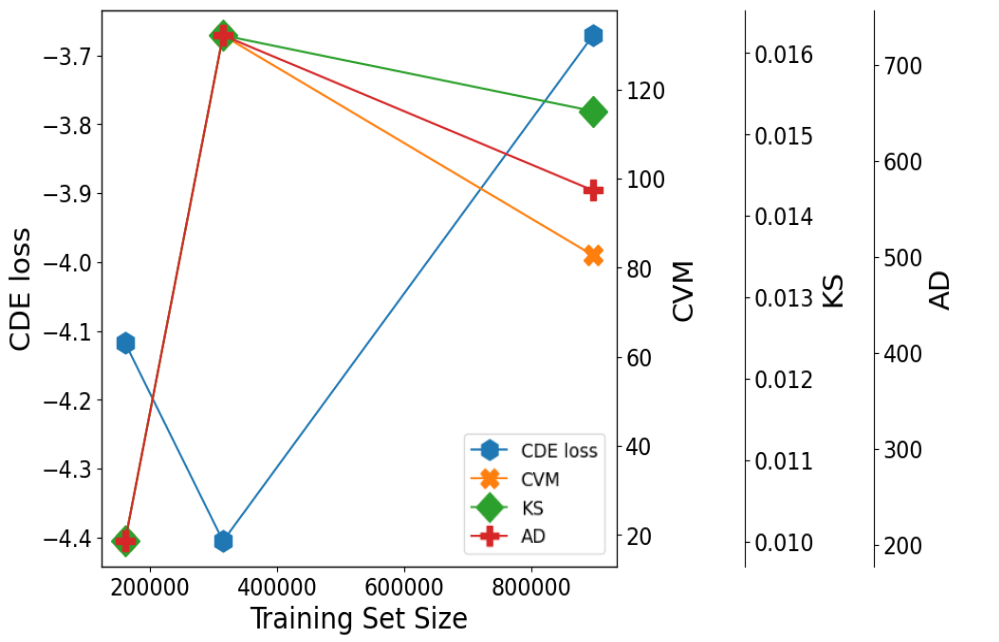}}
    \subfloat{\includegraphics[width=0.45\textwidth]{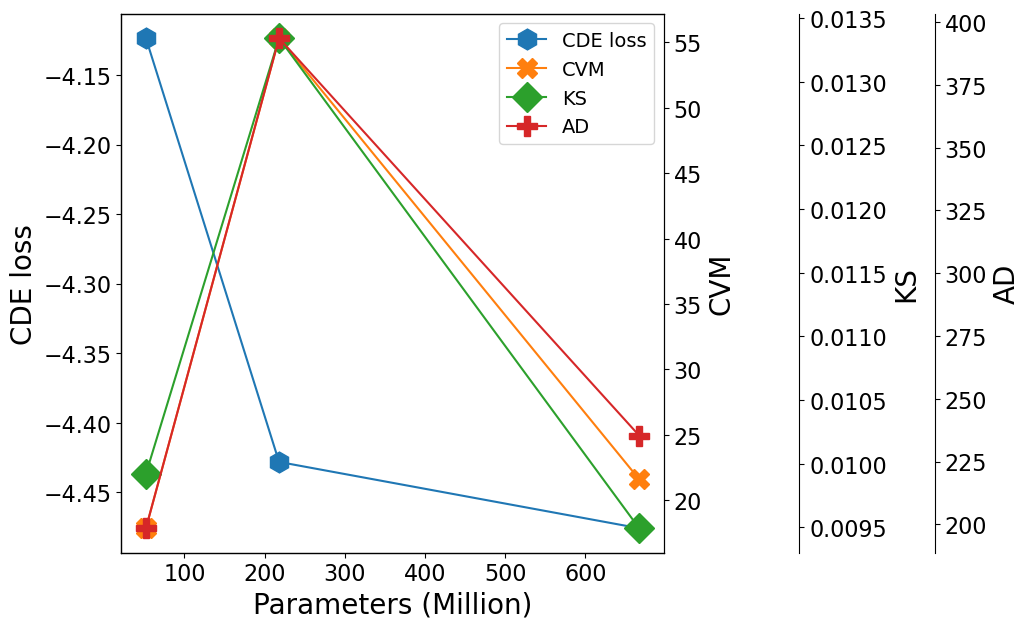}}
    \caption{Scaling laws for redshift PDF estimates.  The top panel shows the effect of increasing the training set size by 2x and 5x, and the bottom panel shows the effect of using larger model backbones, in our case the MViTv2 Base, Large, and Huge models.  Similar to the point estimate metrics, there is no conclusive gain from increasing either the training set size or model size.}
    \label{fig:scaling_laws_training}
\end{figure*}

An investigation of the scaling laws governing our \textsc{DeepDISC} model is shown in Figures \ref{fig:scaling_laws_model} and \ref{fig:scaling_laws_training}.  We train using the Year 5 data and show the effects of independently increasing the model and training set size, quantified by the photo-z metrics used thus far.   Overall, there does not appear to be a strong scaling law for either model size or training set size, in terms of both point estimate and PDF metrics.   This is mostly consistent with \cite{Walmsley24}, as they find that while models pre-trained on ImageNet modestly benefit from increased model size, the largest gains were realized with pre-training in-domain on astronomical images.  Our baseline model appears to have reached a "saturation" of possible gains with increased size or data, indicating other avenues to improve performance may be more fruitful. A full investigation of how \textsc{DeepDISC} scales in different pre-training contexts merits is left to future work.  Here, we are testing whether a simple change to the training set or model size yields noticeable benefits. 

\section{Discussion}
\label{sec:discussion}

\begin{figure*}
    \centering
    \includegraphics[width=\textwidth]{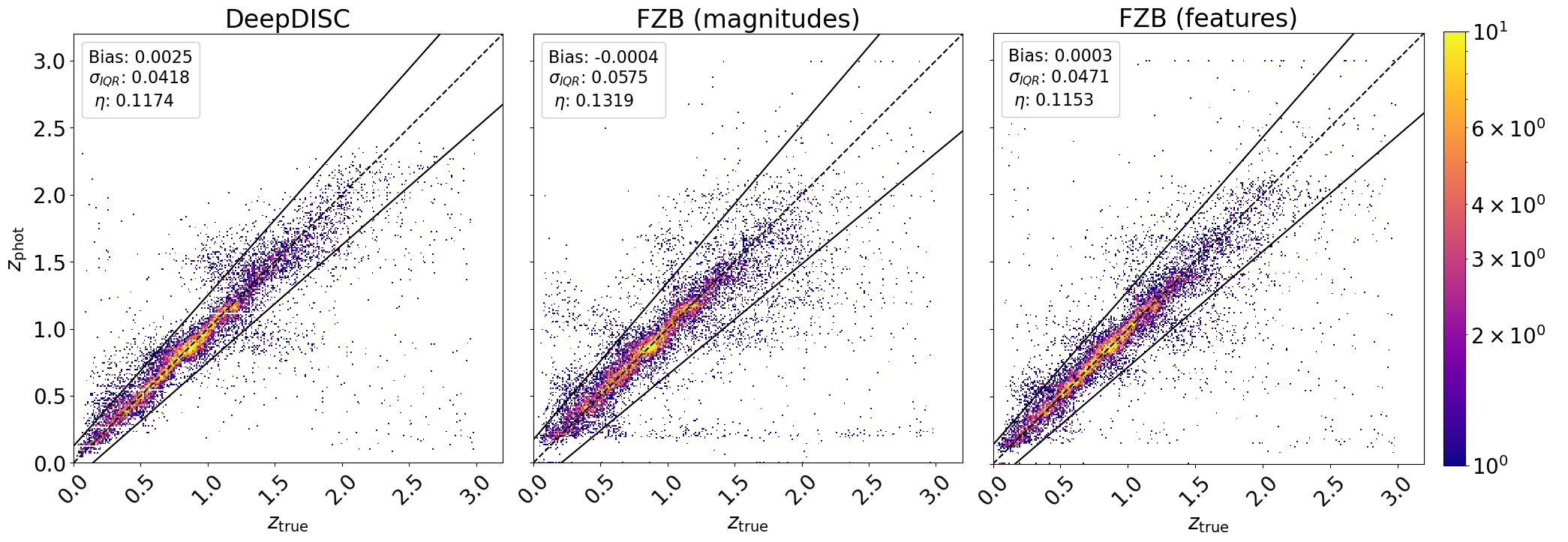}
    \caption{After \textsc{DeepDISC} inference is ran, a small sample is set aside and divided into training and test samples for FZB.  We compare the mode point estimate results on this small test sample.  The dashed line is along $z_{\rm true}= z_{\rm phot}$, and the solid lines define the $3\sigma_{\rm IQR}$ outlier boundary.  \textsc{DeepDISC} is shown on the left, and FZB trained on the small training sample photometry/\textsc{DeepDISC} features is shown in the middle and right, respectively.  Training with \textsc{DeepDISC} features leads to reduced scatter and outlier fractions in the test sample. \textsc{DeepDISC} also produces lower photo-z scatter than FZB trained on the \textsc{DeepDISC}-selected sample.}
    \label{fig:fzb-feature-point-est-comp}
\end{figure*}

\begin{figure*}
    \centering
    \subfloat{\includegraphics[width=0.3\textwidth,height=7cm]{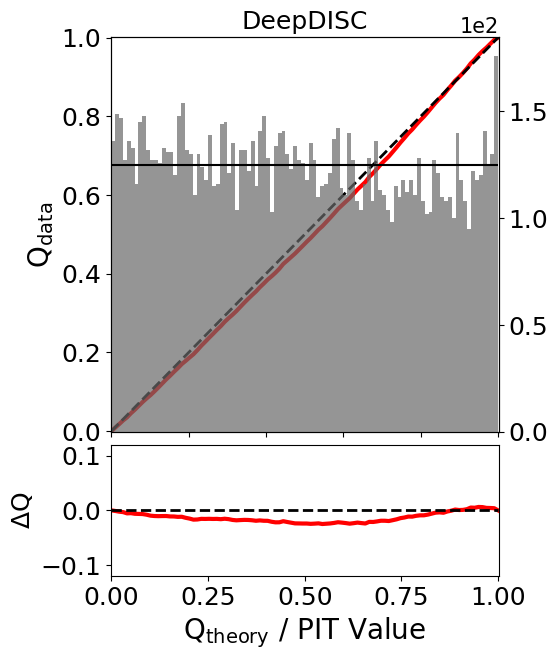}}
    \subfloat{\includegraphics[width=0.3\textwidth,height=7cm]{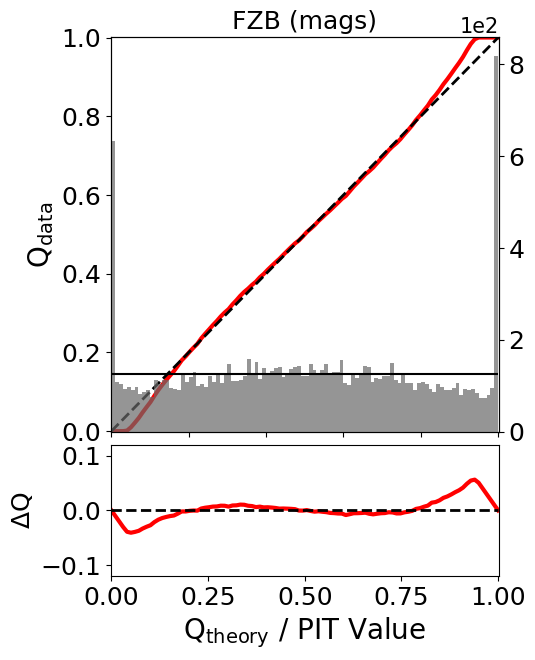}}
    \subfloat{\includegraphics[width=0.3\textwidth,height=7cm]{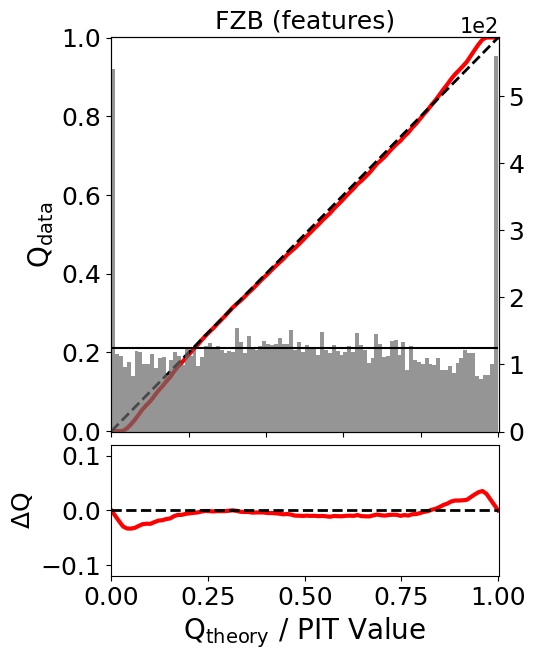}}
    \caption{PIT histograms and QQ plots of the PDF ensembles produced by DeepDISC (left), and FZB trained on photometry (middle) vs \textsc{DeepDISC} features (right).  PIT histograms are shown in grey and correspond to the right y-axis.  The black horizontal line represents a uniform PIT histogram.  The red line plots quantiles of the ensemble PIT values compared to quantiles of an ideal flat distribution and corresponds to the left y-axis.}
    \label{fig:FZB_PIT_hists_comp}

\end{figure*}

\textsc{DeepDISC photo-z} is designed as an end-to-end pipeline for downstream photo-z measurements. Like other computer vision models, \textsc{DeepDISC} skips the need for photometry as an intermediate step and directly produces a photo-z catalog.  The features that the \textsc{DeepDISC} backbone learns during training do not have obvious physical meaning, but do capture relevant information. In the previous sections, we examined DeepDISC performance as a function of object properties and imaging systematics to quantify and understand some of the performance gains relative to traditional catalog-level photo-z algorithms.  \textsc{DeepDISC} appears to be able to handle cases in which traditional algorithms struggle, such as highly blended objects and regions of redshift space in which strong spectral features are not covered by the optical LSST filters.  This performance gain is likely due to the extra information gathered from pixel values and extracted by the backbone network, as opposed to traditional aperture photometry.

We test this hypothesis by decoupling the photo-z estimation from the feature extraction.  To do this, we compare an FZB model trained on photometry with an FZB model trained on the corresponding features produced by a trained \textsc{DeepDISC} model.  The feature tensor output by the backbone is large (256x7x7 for each object) so we first flatten it and run principle component analysis (PCA) to reduce the dimensionality of the features to 12 values per object, matching the size of the photometric catalog (magnitudes + errors for 6 filters).  Some spatial information may be lost due to the flattening, so the results here can be considered a lower bound on the relative performance. We run the trained \textsc{DeepDISC} model on a small subset of images from the main test set. We perform the same catalog-matching and dust reddening corrections described in Section \ref{subsec:cat-codes} and Section \ref{sec:results} to produce a matched catalog of photometry and corresponding \textsc{DeepDISC} features.  We then divide this catalog into a training and test set for FZB.  We train FZB using the same hyperparameters described in Section \ref{subsec:cat-codes}, varying only whether the model uses photometry or \textsc{DeepDISC} features.

The test set point estimates for \textsc{DeepDISC} and both models are shown in Figure \ref{fig:fzb-feature-point-est-comp}. The FZB model trained with \textsc{DeepDISC} features produces less scatter in its point estimates compared to the model trained with photometry (middle versus right panels).  PIT histograms for the models are shown in Figure \ref{fig:FZB_PIT_hists_comp}.  Differences are small, but the model trained with \textsc{DeepDISC} features produces a slightly flatter histogram, and less outliers at the edge of the distribution. Overall, we find that the backbone network is able to extract more information relative to general photo-z estimation than colors/magnitudes derived from aperture photometry.

\subsection{Sample Selection}
In the analysis in Section \ref{sec:results}, the BPZ and FZB methods were trained on catalog photometry matched to the DC2 \texttt{truth} catalog that \textsc{DeepDISC} was trained on.  However, the test set selection is determined by \textsc{DeepDISC} inference matched with the DC2 \texttt{object} catalog, potentially privileging our method as it determines the sample selection that the other codes use for inference.  I.e., the selection effects may make the training data non-ideal for FZB and BPZ in the previous analysis.  This effect is accounted for in the analysis above, as training and test data for FZB is taken from the \textsc{DeepDISC} selection catalog. Comparing the left and middle panels of Figures \ref{fig:fzb-feature-point-est-comp} and \ref{fig:FZB_PIT_hists_comp} shows the relative results when this selection is accounted for.  \textsc{DeepDISC} still produces a lower scatter than FZB, but its bias is slightly higher (although still within LSST science requirements).  The PIT histogram appears slightly flatter for \textsc{DeepDISC}, with less outliers at the edges.  We further discuss the relevance of sample selection in the context of weak lensing below.

\subsection{Weak Lensing}
For studies of weak lensing that require tomographic binning of redshifts to produce $n(z)$ distributions, characterizing major sources of bias in the weak lensing measurement is necessary so that they may be accounted for and mitigated \citep{Mandelbaum18}. Particularly relevant for weak lensing with \textsc{DeepDISC} photo-zs are selection biases due to source detection and photo-z estimation. Source detection may implicitly depend upon galaxy shape through its correlation with lensing.  The photo-z estimation may also be dependent on galaxy shape due to the incorporation of morphological information.  A rigorous investigation of these effects and a full weak lensing analysis is deserving of its own work and outside the current scope here.  However, selection biases have been well-studied in the literature, so there are paths forward.

One way to account for selection effects is through self-calibration, commonly done with the \textsc{metacalibration} algorithm \citep{Metacal,Sheldon17}.  This process involves artificially applying a shear to real galaxy images and measuring the response of some measurement to the shear.  Selection biases can then be quantified and removed from subsequent weak lensing analysis. \textsc{metacalibration} has been applied to galaxy shape estimators \citep{DES_Y3_shape}, photo-z estimators \citep{DES_Y1_WL}, and object detection algorithms \citep{metadetect}.  Future work will adapt \textsc{DeepDISC} to simultaneously estimate redshift and galaxy shapes, providing a single method to detect, deblend, measure and calibrate selection biases.

\subsection{Limitations}
The simulations used in this work include many realistic effects, both due to physical phenomena and instrumental design.  While they encapsulate an enormous effort to represent the universe as close to the ``truth'' as possible, there are still fundamental limitations in our experiment and thus potential differences in how the model will perform on real data compared to the results here.
As described in Section \ref{sec:data}, the morphological model for objects in the DC2 simulations is a simple bulge+disk+knot model.  The size-luminosity relation for disk and bulge components is determined from fitting a half-light radius to the functional form of \cite{Zhang19}, such that parameters for SDSS galaxies are recovered at $z=0$ and sizes decrease with redshift following a sigmoid function.  This relation, along with the simple morphological profiles, cannot capture the true diversity of galaxy shapes and sizes that will be observed by LSST. However, higher fidelity simulations with generative models or real morphological profiles may help to bridge the simulation-reality gap. We are also encouraged by preliminary results on real JWST images (Merz et al., in prep).

In this work, our training sample is completely representative of our test sample, which is a best case scenario for machine learning methods like \textsc{DeepDISC} and FZB.  It is worthwhile to note that in these ideal conditions, \textsc{DeepDISC} outperforms FZB, but the metrics can in some sense be thought of as upper bounds on the performance.  \cite{Moskowitz24} show that underrepresented training samples can severely bias FZB photo-z estimation, and since \textsc{DeepDISC} is also a machine-learning estimator, is it likely also susceptible to this effect.  The solution proposed by \cite{Moskowitz24} is to use data augmentation to help the model learn in underrepresented regimes. For image-based methods such as \textsc{DeepDISC}, simulated images could be generated from augmented catalogs, or the DC2 images themselves could be used to give the networks a more complete picture of the color-redshift space.

\begin{figure}
    \centering
    \includegraphics[width=\columnwidth]{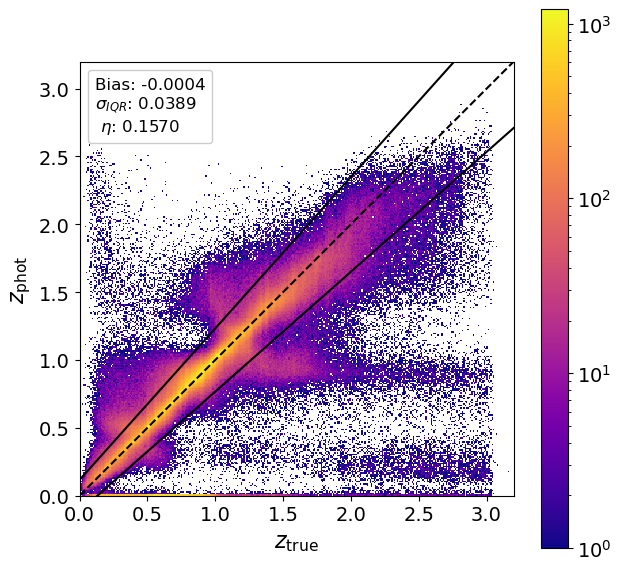}
    \caption{Mode point estimates of the DeepDISC model trained with 5x the amount of training data.  The network is able to better estimate photo-zs above $z_{\textrm{phot}}=2.5$.}
    \label{fig:highz-5x}
\end{figure}

One of the main limitations of \textsc{DeepDISC photo-z} is evident in the lack of PDF modes above $z=2.5$.  While all codes struggle in this high-z regime, it is particularly stark in the \textsc{DeepDISC} scatter plot in Figure \ref{fig:point_estimates}.  Photo-z estimation in this regime is fundamentally a challenge, as the low sample size of objects limits the network's ability to learn relevant features. One potential avenue to improve \textsc{DeepDISC} estimates in the high-z regime is through increasing the amount of high-z data in the training sample.  While we investigated the scaling laws in Section \ref{subsec:scaling} and found no strong improvement of point estimate metrics with increased training set size, those metrics were calculated over the whole sample and did not capture the behavior of the networks in different regimes.  We find that the models trained with 2x and 5x training data are able to provide mode point estimates beyond the $z=2.5$ cutoff (see Figure \ref{fig:highz-5x}). This test increased the number of high-z objects in the training set, but preserved the overall $n(z)$.  In future work, targeted data augmentation like that of \cite{Moskowitz24} could be implemented in order to specifically increase $n(z>2.5)$. The models are very sensitive to regimes with low representation, so added data does make a difference.  Different pre-training schemes and architectures may also lead to improved high-z performance and will be explored in follow-up work.

\section{Conclusions}
\label{sec:conclusions}

Photometric redshift estimation will be of major importance in upcoming Wide-Fast-Deep surveys for precision cosmology, studies of galaxy evolution, and more.  A multitude of different photo-z codes have been developed under different frameworks.  While codes that use catalog level photometry remain popular, a growing focus has been applied to image-based codes that use deep learning to map pixel level inputs to redshift estimations.  In this work, we present \textsc{DeepDISC} photo-z, an image-based photo-z estimator built on the \textsc{DeepDISC} framework \citep{Merz2023}.  \textsc{DeepDISC} utilizes object detection networks to simultaneously find objects and make predictions in an image. This provides an advantage over image-based codes that typically assume each input image has a single object located at the center.  \textsc{DeepDISC} makes no assumptions about the spatial location or the number of objects, lending itself to application on massive datasets of raw images.  Another advantage of \textsc{DeepDISC} is its ability to be tuned \textit{after} training.  For instance, it can be applied to images of variable size.  

In this work we train on image cutouts with a size of 525 pixels square and apply the trained network to cutouts of size 1050 pixels square to demonstrate the ability of the network to generalize to large scenes.  Additional hyperparameters can be tuned after training to increase the detection sensitivity. In order to validate \textsc{DeepDISC photo-z}, we conduct a controlled experiment using simulated LSST DESC DC2 data.  We compare \textsc{DeepDISC} photo-z to the catalog-level codes \textsc{Bayesian Photometric Redshifts}{,} which is template-based and \textsc{FlexZBoost}, which uses machine learning. 

Our model is trained on 4 NVIDIA V100 GPUs for 6 hours and estimates 3 million photo-zs on a single V100 GPU in 40 minutes.  Though our model is computationally expensive to train compared to the catalog-based codes used in this study, it is fairly competitive in estimating photo-zs on a large test set.  When parallelized across 16 CPUs, BPZ estimation takes roughly 0.001 seconds per object, and FZB estimation takes 0.0002 seconds per object.  Considering \textsc{DeepDISC} produces a catalog roughly 3 times the size as the input catalog used for BPZ and FZB estimation, our model takes roughly 0.0008 seconds per object to produce a photo-z PDF when ran on a single GPU. \textsc{DeepDISC} can be parallelized to estimate photo-zs across multiple GPUs, but more CPUs than GPUs are generally available for researchers to use for parallelization, so FZB and BPZ have an advantage in scaling on large datasets.  However, we note that \textsc{DeepDISC} runtime per image does not scale strongly with the size of the image, and large enough GPUs could potentially handle a full-sized CCD image (4k pixels square for LSST).

In order to conduct a fair comparison of the codes, we supply the other codes the object catalogs, which have been reduced from the input images.  This ensures all methods are trained and applied to data that accounts for imaging systematics such as blending.  We find that \textsc{DeepDISC} outperforms the catalog-based codes in almost all metrics on the test set, including metrics designed to evaluate the quality of point estimates and uncertainty predictions.  We find that \textsc{DeepDISC} photo-z tends to under-predict high redshifts and does not produce mode point estimates above $z\sim2.5$, likely due to the low representation of these objects in the training.  However, this may be ameliorated with additional high-z data augmentation.

We compare the quality of the predicted redshift PDFs of all codes by examining the PIT histograms and probabilistic metrics. \textsc{DeepDISC} produces the empirical PIT histogram closest to the ideal theoretical uniform distribution, shown in the visualization and statistical measurements. We find that we are able to recover a majority of photo-z outliers (i.e., they are no longer considered outliers) if $z_{\rm phot}$ is taken be the secondary peak of the PDF, indicating that meaningful degeneracies are learned by the network.

Bias, scatter and outlier fraction are examined as a function of systematics that affect the imaging and thus could potentially affect the photo-z estimates.  Overall, we find no strong dependence on dust extinction, or PSF FWHM.  However, we find that \textsc{DeepDISC} is much more robust at estimating photo-zs for blended sources, indicating that it is able to learn rich information at the pixel-level.  We do not fully characterize the effects of object detection and unrecognized blends in this study. Unrecognized blends arise when only a single object is detected in a blend consisting of multiple sources, and can make up 15-20\% of total objects \citep{Dawson16,Troxel23}.  In this work, we have chosen to compare our DeepDISC photo-z catalog to BPZ and FZB by cross-matching DeepDISC detections and the DC2 \texttt{object} catalog produced by LSST science pipelines.  Thus, some of the objects in our test set will be unrecognized blends due to non-detections in the LSST science pipeline.  Follow-up work to isolate this effect and other potential differences due to the different detection methods will include producing a photometric catalog by running forced photometry on the images using DeepDISC detections, and further removing samples of unrecognized blends based on truth catalogs.  

We test \textsc{DeepDISC} photo-zs using imaging created from 1 year worth of observations and 5 years worth of observations.  With the signal-to-noise increased by a factor of $\sim$2.24 (SNR $\propto \sqrt{{\rm{time}}}$), we see a roughly similar factor of decrease in scatter.  We also see the network estimate more high-z objects with the 5 year observations, as more signal is collected for these faint sources.  

In addition to the tests above, we examine possible scaling laws that could govern our model.  The backbone of the network is a MViTv2 network, a vision transformer which is known for being data hungry \citep{zhai2022scaling}. We investigate if our models (pre-trained on ImageNet) have notable performance gains when the data set size or model size is increased.  We find no strong relation for any metric, point or probabilistic, as a function of model size or training set size.  Scaling studies often find power law dependencies, so it is possible we are already in the saturated regime as we do not see any notable gains.  \cite{Walmsley24} found that while pre-training on ImageNet data can improve downstream galaxy morphology tasks and does scale with data set and model size, most performance gains come from additional pre-training on galaxy images. Investigation into this effect for \textsc{DeepDISC} will be a promising avenue of future work, especially as pre-training using galaxy images has been shown to improve the detection of galactic features through a Faster-RCNN network \citep{Popp24}. Pre-training on simulated images (such as those used in this study) may help with training the model on real data, which is incomplete in redshift and color-magnitude space due to spectroscopic selection effects.  Training sample augmentation has been shown to mitigate this bias and it would be worthwhile to explore whether \textsc{DeepDISC} would benefit from this method.  \cite{Moskowitz24} augmented a photometric catalog to produce a new training set by using a normalizing flow, and such a catalog could be used to simulate new images to produce an augmented training set for \textsc{DeepDISC}. 

We have demonstrated that \textsc{DeepDISC} is able to learn more information relevant for photo-z estimation than catalog photometry can provide.  This may help in redshift regimes where strong spectral features are not visible through the available filters, although the data used in this study is idealized in some ways compared to real morphological profiles and SEDs. We will continue to test our model on real data for practical implementations. Individual photo-z PDFs are useful for a variety of science cases, but for future applications that use $n(z)$ such as weak lensing, thorough investigation of the selection biases of \textsc{DeepDISC} will be done to properly calibrate measurements. \textsc{DeepDISC} has shown promising performance on simulated LSST data, which we aim to continue to improve and vet as we quickly approach Rubin first light.

\section*{Acknowledgements}

G.M. led the initial DeepDISC codebase development, data curation, analysis, and paper writing. X.L. advised G.M., planned tests and analyses of the model and contributed to writing the paper.  S.S. provided expertise and code regarding the use of catalog-based photo-z methods, suggestions for analyses, and paper edits.  A.M. contributed to the initial formalization of rail\_deepdisc, provided expertise regarding RAIL, and paper edits.  T.Z. contributed to the DeepDISC codebase, reviewed first draft figures and provided paper edits and suggestions. D.O. co-led the development of the codebases with G.M.  D.B., M.D., J.K and O.L. developed both DeepDISC and rail\_deepdisc codebases.  C.B. led the initial work that inspired the creation of DeepDISC and contributed paper edits.  Y.L. and Y.E. contributed to the codebase.

This paper has undergone an internal review by the
LSST DESC, and we thank the internal reviewers, Eric Gawiser and Ismael Mendoza, for thoughtful feedback that improved the paper.
We thank Athol Kemball, Matias Carrasco Kind, Yuxiong Wang and Jeff Newman for illuminating discussions, and the anonymous referees for helpful comments. 
We thank Shirui Luo, Dawei Mu, and Volodymyr Kindratenko at the National Center for Supercomputing Applications (NCSA) for their assistance with the GPU cluster used in this work. 
G.M., X.L., Y.E. and Y.L. acknowledge support from the NCSA Faculty Fellowship, NCSA Student Pushing INnovation internship program, LSST-DA through grant 2023-SFF-LFI-03-Liu, NSF grant AST-2308174, and NASA grant 80NSSC24K0219. 
G.M. thanks the LSST-DA Data Science Fellowship Program, which is funded by LSST-DA, the Brinson Foundation, and the Moore Foundation; his participation in the program has benefited this work.
This work was conducted as part of a LINCC Frameworks Incubator. LINCC Frameworks is supported by Schmidt Sciences.
%
AIM, TZ, DO, OL, DB, JK, and MD are supported by Schmidt Sciences.
C.J.B. is supported by an NSF Astronomy and Astrophysics Postdoctoral Fellowship under award AST-2303803. This material is based upon work supported by the National Science Foundation under Award No. 2303803. This research award is partially funded by a generous gift of Charles Simonyi to the NSF Division of Astronomical Sciences. The award is made in recognition of significant contributions to Rubin Observatory’s Legacy Survey of Space and Time. 
This work utilizes resources supported by the National Science Foundation's Major Research Instrumentation program, grant \#1725729, as well as the University of Illinois at Urbana-Champaign. 

The DESC acknowledges ongoing support from the Institut National de 
Physique Nucl\'eaire et de Physique des Particules in France; the 
Science \& Technology Facilities Council in the United Kingdom; and the
Department of Energy, the National Science Foundation, and the LSST 
Corporation in the United States.  DESC uses resources of the IN2P3 
Computing Center (CC-IN2P3--Lyon/Villeurbanne - France) funded by the 
Centre National de la Recherche Scientifique; the National Energy 
Research Scientific Computing Center, a DOE Office of Science User 
Facility supported by the Office of Science of the U.S.\ Department of
Energy under Contract No.\ DE-AC02-05CH11231; STFC DiRAC HPC Facilities, 
funded by UK BEIS National E-infrastructure capital grants; and the UK 
particle physics grid, supported by the GridPP Collaboration.  This 
work was performed in part under DOE Contract DE-AC02-76SF00515.

We acknowledge use of Matplotlib \citep{Hunter2007}, a community-developed Python library for plotting. This research made use of Astropy,\footnote{\href{http://www.astropy.org}{http://www.astropy.org}} a community-developed core Python package for Astronomy \citep{astropy:2013, astropy:2018}. This research has made use of NASA's Astrophysics Data System.


This paper makes use of software developed for the Large Synoptic Survey Telescope. We thank the LSST Project for making their code available as free software at  http://dm.lsst.org



This research has made use of the NASA/IPAC Infrared Science Archive, which is funded by the National Aeronautics and Space Administration and operated by the California Institute of Technology.




\bibliographystyle{mnras}
\bibliography{ref} 



\appendix

\section{Smoothed Images}
\label{app:blurred}
\begin{figure*}
    \centering
    \includegraphics[width=0.7\textwidth]{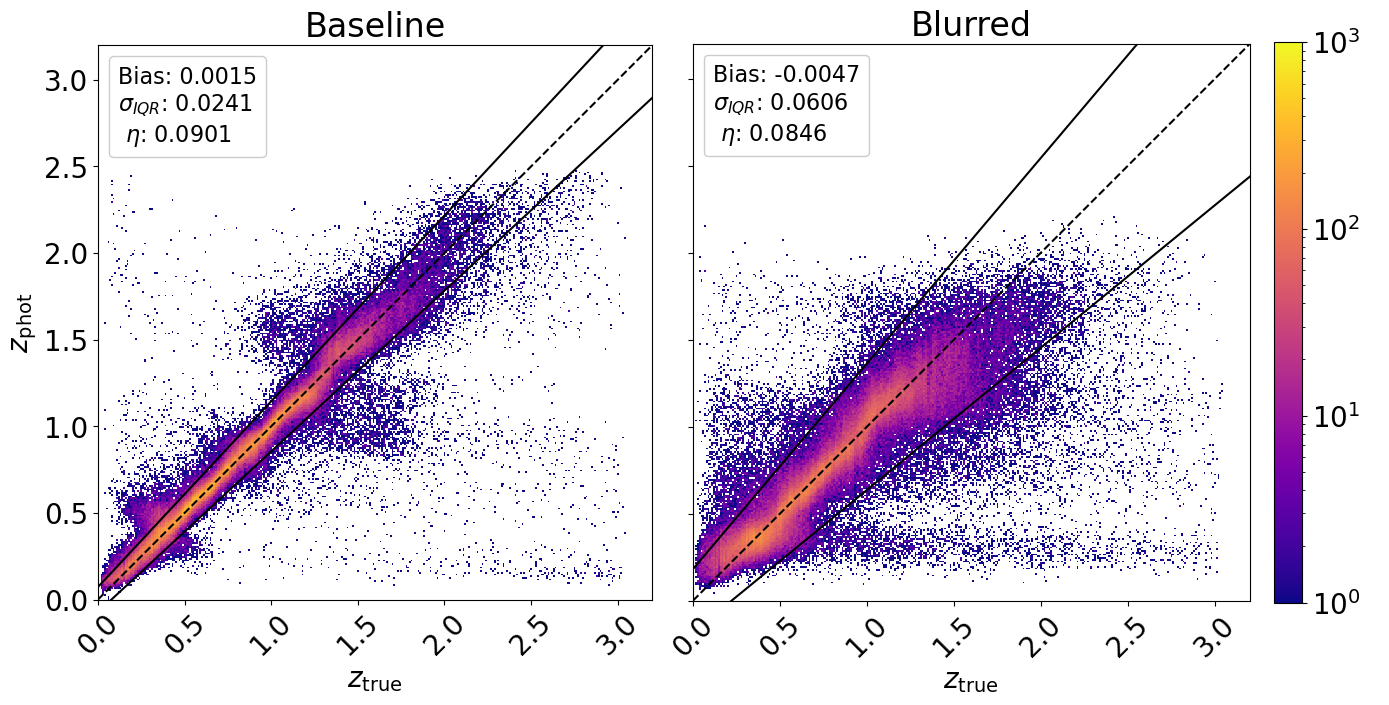}
    \caption{The baseline \textsc{DeepDISC} model (left) compared to a model trained on blurred images (right).  The model performs significantly worse on the blurred images, indicating that \textsc{DeepDISC} utilizes pixel-level information to extract more information about the sources than just colors and magnitudes.  The black, dashed line follows the $z_{\rm true}= z_{\rm phot}$ line, and solid lines define the 3$\sigma_{\rm IQR}$ outlier boundary.}
    \label{fig:blur-test}
\end{figure*}

Image-based photo-z estimation has been shown to be competitive with and often outperform traditional catalog-based photo-z estimation.  Pixels contain more information about sources than aperture photometry, such as color gradients and detailed morphological features.  Interestingly, previous studies have found that adding morphological measurements to catalog-level photo-z codes do not significantly improve the photo-z estimations \citep{Soo18, Wilson20}. However, image-based estimators remain extremely competitive.  Extracting features directly from the pixels gives deep neural networks an access to more information over methods that use pre-computed features, even if those features include morphology.  We show the importance of the pixel-level information by comparing \textsc{DeepDISC} photo-zs to the same model trained with images convolved with a Gaussian kernel of 25 pixels.  This blurring removes pixel-level morphological information but preserves overall flux and color for a given source.  We then take the intersection of objects detected by each model, and compare photo-zs in Figure \ref{fig:blur-test}. The model trained and evaluated on blurred images performs worse than the baseline, indicating that morphological information contained at the pixel level is beneficial for the network.  This is consistent with \cite{Schuldt21}, who find that a convolutional neural network trained on multi-band PSF images scaled to object magnitudes performs worse than a network trained on the original images.

\section{Stellar Contamination}

\label{app:stars}
Towards the detection limit of a survey, stars and galaxies become hard to distinguish because faint galaxies tend to appear smaller and approach the resolution limit of the telescope.  Thus, completeness and purity of star and galaxy samples at faint magnitudes tends to drop \citep{Bosch18}.  \cite{DIsanto18} test the performance of a deep neural network photo-z algorithm on a sample that includes stars, galaxies, and quasars, eliminating the need for pre-classification.  In our case, \textsc{DeepDISC} performs classification along with all other tasks simultaneously.  These amount to object detection, segmentation, classification, and photo-z estimation.  Our results shown in the main body of this work have been produced from a pure galaxy sample using a network with the photo-z head trained only on galaxies (but all other heads trained with stars+galaxies). Here, we examine the effect of including stars in our training and evaluation samples.

The classification head of the network uses a fully connected and a softmax layer to output the probability of a detected object belonging to each class (star/galaxy).  The completeness and purity of our classifier compared to the LSST Science Pipelines extendedness metric is shown in Table \ref{tab:class_pr}.  \textsc{DeepDISC} obtains a higher galaxy completeness, as well as stellar purity.  The extendedness metric yields a slightly higher stellar completeness.
\newline

\begin{table}
    \centering
    \begin{tabular}{ccc}
    \hline
    \hline
         \textsc{DeepDISC}  &  Completeness & Purity \\
         Galaxy &  0.998 & 0.989 \\
         Star   &  0.785 & 0.961 \\
         \hline
         LSST  &   &  \\
         Galaxy &  0.950 & 0.989 \\
         Star   &  0.795 & 0.446 \\
    \hline
    \end{tabular}
    \caption{Star/Galaxy classification for \textsc{DeepDISC} and LSST extendedness}
    \label{tab:class_pr}
\end{table}

\begin{table}
    \centering
    \begin{tabular}{cccc}
        \hline
        \hline
         & bias & $\sigma_{\rm IQR}$ & $\eta$  \\ 
         \hline
         \textit{z-pure} (pure eval) & 0.0007  & 0.0410 &  0.1183  \\
         \textit{z-pure} (contam eval) & 0.0014  & 0.0422 &  0.1276  \\
         \textit{z-all} (contam eval)  & 0.0002  & 0.0422 &  0.1279  \\
        \hline
    \end{tabular}
    \centering
    \label{tab:pure-contam}
    \caption{Comparing \textsc{DeepDISC} photo-zs when only training and evaluating on a pure galaxy set (first row) vs training with stars and selecting the evaluation set based on the network classification (bottom row).  The network that has been trained to handle stars }
\end{table}

We define a model trained and evaluated only on galaxies as \textit{z-pure}, and a model trained with stars included in the redshift branch as \textit{z-all}.  In Table \ref{tab:pure-contam} we compare point estimate metrics of \textit{z-all} and \textit{z-pure} evaluated on galaxies selected by the \textsc{DeepDISC} classifier, i.e. with some small amount of stellar contamination. Comparing rows one and two shows the effect of stellar contamination.  Although we filter stars with our classifier in row 2, some contamination remains.  This leads to an increase in bias, scatter, and outlier fraction.  Comparing rows two and three shows the effect of adding stellar redshifts ($z=0$) to the photo-z head during training.  Adding stars to the redshift head does not significantly affect the results.  It appears that with some level of stellar contamination in the evaluation set, \textsc{DeepDISC} photo-zs do not necessarily benefit from including stars in the training.

\label{lastpage}
\end{document}